%% file: Holograph_MIMO.tex
\documentclass[journal]{IEEEtran}
\usepackage{tikz}
\usetikzlibrary{arrows,positioning,fit,shapes,calc,decorations.pathreplacing,matrix,patterns}
\usepackage{amsmath}
\usepackage[caption=false,font=footnotesize,labelfont=rm,textfont=rm]{subfig}
\usepackage{stfloats}
\usepackage{lineno,hyperref}
\usepackage{xcolor}
\usepackage{extarrows}
\usepackage{algorithm}
\usepackage{algorithmic}
\usepackage{amssymb}
\usepackage{graphicx}
\usepackage[justification=centering]{caption}
\usepackage{cite}

\usepackage{stfloats}
\usepackage{amsthm}
\usepackage{geometry}
\usepackage{booktabs}
\usepackage{cuted}

\input{mydefinitions1.tex}

\begin{document}
	\title{Neural Network-Assisted Hybrid Model Based Message Passing for Parametric Holographic MIMO Near Field Channel Estimation} 
	\author{Zhengdao~Yuan, Yabo Guo, Dawei Gao, Qinghua Guo, \IEEEmembership{Senior Member, IEEE}, Zhongyong Wang, Chongwen Huang, Ming Jin and Kai-Kit Wong, \IEEEmembership {Fellow, IEEE} \vspace{-1.5em}
		\thanks{Corresponding authors: Qinghua Guo and Zhongyong Wang.}
		\thanks{Z. Yuan is with the Artificial Intelligence Technology Engineering Research Center, Henan Open University,
			Zhengzhou 450002, China. He was with the School of Electrical, Computer and Telecommunications Engineering, University of Wollongong, Wollongong, NSW 2522, Australia (e-mail: yuan\_zhengdao@foxmail.com).}
		\thanks{Y. Guo and Z. Wang are with the School of Information Engineering, Zhengzhou University, Zhengzhou 450002, China. Y. Guo is also with the School of Electrical, Computer and Telecommunications Engineering, University of Wollongong, Wollongong, NSW 2522, Australia  (e-mail: ieybguo@163.com, zywangzzu@gmail.com).}
		\thanks{D. Gao is with the Hangzhou Institute of Technology, Xidian University, Hangzhou 311200, China.
			He was with with the School of Electrical, Computer and Telecommunications Engineering, University of Wollongong, NSW 2522, Australia
			(e-mail: gaodawei@xidian.edu.cn).}
		\thanks{Q. Guo is with the School of Electrical, Computer and Telecommunications Engineering, University of Wollongong, Wollongong, NSW 2522, Australia  (e-mail: qguo@uow.edu.au).}
		\thanks{C. Huang is with the College of Information Science and Electronic Engineering, Zhejiang University, Hangzhou 310007, China, and Zhejiang Provincial Key Lab of Information Processing, Communication and Networking (IPCAN), Hangzhou 310007, China, and the International Joint Innovation Center, Zhejiang University, Haining 314400, China (e-mail: chongwenhuang@zju.edu.cn).}
		\thanks{M. Jin is with the Faculty of Electrical Engineering and Computer Science, Ningbo University, Ningbo 315211, China (e-mail: jinming@nbu.edu.cn)} 
		\thanks{K. K. Wong is affiliated with the Department of Electronic and Electrical Engineering, University College London, Torrington Place, WC1E 7JE, United Kingdom and he is also affiliated with the Department of Electronic Engineering, Kyung Hee University, Yongin-si, Gyeonggi-do 17104, Korea (e-mail: kai-kit.wong@ucl.ac.uk).} 
	}

	\maketitle
	
	\begin{abstract}
		Holographic multiple-input and multiple-output (HMIMO) is a promising technology with the potential to achieve high energy and spectral efficiencies, enhance system capacity and diversity, etc.
		In this work, we address the challenge of HMIMO near field (NF) channel estimation, which is complicated by the intricate model introduced by the dyadic Green’s function. 
		Despite its complexity, the channel model is governed by a limited set of parameters. This makes parametric channel estimation highly attractive, offering substantial performance enhancements and enabling the extraction of valuable sensing parameters, such as user locations, which are particularly beneficial in mobile networks. However, the relationship between these parameters and channel gains is nonlinear and compounded by integration, making the estimation a formidable task.
		To tackle this problem,  we propose a novel neural network (NN) assisted hybrid method. 
		With the assistance of NNs, we first 
		develop a novel hybrid channel model with a significantly simplified expression compared to the original one, thereby enabling parametric channel estimation. Using the readily available training data derived from the original channel model, the NNs in the hybrid channel model can be effectively trained offline. Then, building upon this hybrid channel model, we formulate the parametric channel estimation problem with a probabilistic framework and design a factor graph representation for Bayesian estimation. Leveraging the factor graph representation and unitary approximate message passing (UAMP), we develop an effective message passing-based Bayesian channel estimation algorithm. Extensive simulations demonstrate the superior performance of the proposed method.
		
	\end{abstract}
	
	
	\section{Introduction}
	
	\IEEEPARstart {H}{olographic} multiple-input multiple-output (HMIMO) fulfills the deployment of extremely large and near spatial continuous surfaces within a compact space, harnessing the potential of electromagnetic (EM) channels. Recognized as a key enabling technology in future wireless communications, particularly in light of its potential integration into 6G networks, HMIMO holds substantial benefits in achieving high spectral and energy efficiencies, improves the system capacity and diversity, enhances massive connectivity, etc \cite{An2023CL2, An2023CL1,Wei2022TSP,Wei2023TWC, Adhikary2024Trans,Huang2020WC,Luca2023TWC,Deng2021TVT} 

Recently, there has been a notable surge in research on HMIMO, 
leading to a multitude of investigations into its diverse applications in communication systems. Assuming perfect channel state information (CSI), studies such as those in \cite{An2023CL2} and \cite{Deng2021TVT} have delved into beamforming designs employing HMIMO for wireless communications. 
Research efforts like those in \cite{Wei2022TSP} and \cite{Wei2023TWC} have concentrated on tailored precoding designs for HMIMO systems, while others, such as those in \cite{Amico2022TSP} and \cite{DengJSAC2022}, have addressed the holographic positioning problem. 
Additionally, integrated holographic sensing and communications have been explored in works such as \cite{Adhikary2024Trans} and \cite{Zhang2022JSAC}. Moreover, wireless power transfer has been extended to HMIMO systems \cite{Huang2020WC}. Further exploration into wavenumber-division multiplexing within line-of-sight HMIMO communications has been undertaken in \cite{Luca2023TWC}. To leverage the full potential of HMIMO, efficient acquisition of accurate CSI is indispensable.


Some channel estimation methods have bee proposed for HMIMO communications. In \cite{Demir2022WCL}, considering both non-isotropic scattering and directive antennas, a channel model containing angle information for HMIMO was developed, and a novel channel estimation scheme that exploits the rank deficiency induced by the array geometry was proposed, where it does not require the exact channel statistics. With proper approximations to the channel covariance matrix, the work in \cite{Luca2023arXiv} designed a low-complexity scheme to perform HMIMO channel estimation 
which can achieve the same performance as the optimal minimum mean square error (MMSE) estimator. 
A low-complexity Bayes-optimal channel estimator operating in unknown EM environments was proposed in \cite{yu2023bayes} for HMIMO systems, which has no requirements on priors or supervision that is not feasible in practical deployment. The work in \cite{yu2023learning} proposed a self-supervised machine learning channel estimation algorithm, which is designed to operate under more relaxed prior information. 
The work in \cite{chen2023angular} proposed decomposition and compressed deconstruction-based variational Bayesian inference to estimate azimuth and elevation angles, distance parameters, and sparse channels. 
In \cite{Ghermezcheshmeh2023access}, leveraging the specific structure of the radiated beams generated by the continuous surface, a method based on a parametric physical channel model was proposed to estimate the channel of line-of-sight dominated HMIMO in millimeter or THz bands.  
The aforementioned channel estimation methods are based on some simplified channel models with far-field assumption,  
which either break down at the near-field (NF) region or cannot capture the full-polarized information of EM fields \cite{Yuan2022Antennas,Yuan2022Conference}. {However, due to the large aperture of HMIMO surface and the use of high frequency band, which lead to a large Rayleigh distance, there is a need to consider HMIMO communications in NF,} necessitating the use of the Dyadic Green's function to characterize the channels with intractable integration and nonlinearity \cite{Guo2024Trans,Wei2023TWC}. To the best of our knowledge, HMIMO NF channel estimation has not been well addressed in the literature. 

Despite the complexity of the channel model introduced by the dyadic Green's function, it is governed by a limited set of parameters. Compared to direct channel estimation methods that directly estimate a huge number of channel coefficients, parametric channel estimation is expected to achieve substantial performance enhancement. 
In addition, parametric channel estimation enables the extraction of valuable parameters, such as user locations, facilitating sensing in the networks.
However, the relationship between the parameters and channel coefficients exhibits a convoluted nonlinearity compounded by integration, rendering parametric channel estimation a formidable task.
To tackle this challenge, we propose a novel neural network (NN)-assisted hybrid approach. 
With the assistance of NN, We first develop a novel hybrid HMIMO channel model, featuring a significantly simplified expression compared to the original one. This hybrid channel model enables parametric channel estimation. Using readily available training data derived from the original channel model, the NNs in the hybrid channel model can be effectively trained, which can be carried out offline. 
Subsequently, building upon this hybrid channel model,
we formulate the parametric channel estimation problem in a probabilistic form for Bayesian estimation. With a factor graph representation of the parametric channel estimation problem and leveraging the unitary approximate message passing (UAMP) \cite{guo2015approximate,yuan2020approximate, Luo2021Tsp} 
an effective message passing-based Bayesian channel estimation algorithm is developed. Extensive simulation results are provided to demonstrate the superior performance of the proposed method. 
The main contributions of this work are summarized as follows:
\begin{itemize}
\item{To the best of our knowledge, this is the first work on parametric channel estimation of HMIMO NF channels that are characterized using the Dyadic Green's function.}
\item{Considering that HMIMO NF channels are governed by a small set of parameters, we estimate the parameters and subsequently reconstruct the channels, rather than directly estimating a large number of channel coefficients. This parametric approach leads to superior performance as the number of variables to be estimated is drastically reduced, and it also facilitates the sensing function in the system.}
\item{To deal with the intractable Dyadic Green's function based channel model, we propose a NN-assisted hybrid channel model, 
which can be well-trained offline. The NN-assisted hybrid channel model plays a crucial rule in designing a practical HMIMO channel estimation algorithm.}
\item{Building on the hybrid channel model,  we formulate the parametric channel estimation problem into a probabilistic form and develop an effective message passing-based Bayesian channel estimation algorithm, leveraging UAMP.}
\item{Extensive simulation results demonstrate the superior performance of the proposed method.}
\item{Although this work focuses on HMIMO NF channel estimation, the hybrid model approach can be used to tackle a generic signal estimation problem involving a system transfer function, which is intractable using conventional approaches.}
\end{itemize}

{The remainder of this paper is organized as follows. In Section II, we introduce the signal model and problem formulation of HMIMO NF channel estimation. In Section III, a new hybrid channel model is proposed and the channel estimation problem is reformulated, and a factor graph representation is developed. Leveraging UAMP, a message passing algorithm is developed in Section IV. Numerical results are provided in Section V, followed by conclusions in Section VI.}

\textit{{Notations}}: 
Boldface lower-case and upper-case letters denote vectors and matrices, respectively. 
A Gaussian distribution of $\bx$ with mean {$\hat\bx$} and covariance matrix $\bV$ is represented by $\CN(\bx;{\hat \bx},\bV)$.
The relation $f(x)=cg(x)$ for some positive constant $c$ is written as $f(x)\propto g(x)$, and $diag(\ba)$ returns a diagonal matrix with $\ba$ on its diagonal. We use $\bA\cdot\bB$ and $\bA\cdot/\bB$ to denote the element-wise product and division between $\bA$ and $\bB$, respectively. 
The notation $|\bA|^{.2}$ denotes an element-wise magnitude squared operation for $\bA$, and $||\bA||$ is the Frobenius norm of $\bA$. 
We use $\textbf{1}$, $\textbf{0}$ and $\bI$ to denote an all-one matrix, all-zero matrix and identity matrix with a proper size, respectively. {The notation $x \sim \text{U}[a,b]$ denotes that $x$ has a uniform distribution over $a$ and $b$.}

\section{Channel Model and Problem Formulation for HMIMO NF Channel Estimation}

\subsection{Channel Model Using the Dyadic Green's Function}
\begin{figure}[!t]
\centering
\includegraphics[width=0.5\textwidth]{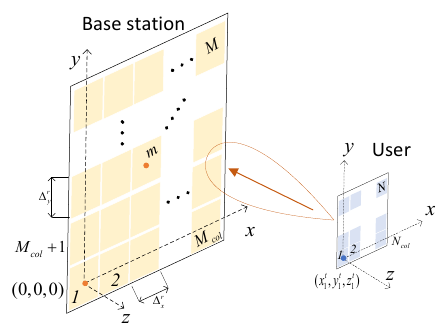}
\centering
\caption{Illustration of antenna patches and their coordinates.}
\label{fig:Patch}
\end{figure}

We consider an HMIMO communication system shown in Fig. \ref{fig:Patch}, where the receiver (base station) and transmitter (user) are equipped with holographic surfaces, comprising ${M=M_{row} \times M_{col}}$ and ${N=N_{row} \times N_{col}}$ patch antennas, respectively. Each patch can transmit or receive signals in three polarizations \cite{Wei2023TWC}. Each transmit patch has a size of $\Delta_x^t\times \Delta_y^t$, and each receive path has a size of $\Delta_x^r\times \Delta_y^r$, where $\Delta_x^r, \Delta_y^r, \Delta_x^t$ and  $\Delta_y^t$ denote the  horizontal and vertical dimensions of receive and transmit patches. As shown in Fig. \ref{fig:Patch}, we number the transmit and receive patches row by row. Assume that the receive surface lays in the $xy$ plane and the center of the first receive patch is located at the origin of the coordinate system. 
The transmit surface is in parallel with the receive surface and the coordinate of the first transmit patch is denoted as $\br_1^t=(x_1^t, y_1^t, z_1^t)$. 
Denote the $m$-th receive patch and the $n$-th transmit patch as $S_m^r$ 
and $S_n^t$, 
respectively. The coordinate vectors of the $m$-th receive and the $n$-th transmit patch centers are $\br_m^r=[x_m^r,y_m^r,z_m^r]^T$ and $\br_n^t=[x_n^t,y_n^t,z_n^t]^T$, respectively. Then the coordinates of the patch centers have the following relationship 
\begin{eqnarray}
&&\!\!\!\!\!\!\!\!\!x_m^r=(c_m^r-1)\Delta_x^r, \ \  
y_m^r=(l_m^r-1)\Delta_y^r, \ \ z_m^r=0,	 \nonumber  \\
&&\!\!\!\!\!\!\!\!\!x_n^t={x_1^t}+(c_n^t-1)\Delta_x^t, 
y_n^t={y_1^t}+(l_n^t-1)\Delta_y^t, z_n^t=z_1^t \label{eq:coord_xmn}
\end{eqnarray}
where $c_m^r=\text{mod}(m,M_{col})$, $l_m^r=\text{ceil}(m/M_{col})$, $c_n^t=\text{mod}(n,N_{col})$, $l_n^t=\text{ceil}(n/N_{col})$, 
$\text{mod}(\cdot)$ denotes the modulo operation and $\text{ceil}(\cdot)$ returns an smallest integer that is greater than or equal to the number in the parentheses. 


{We use  $\bJ(\br^t)$ to denote the current generated at location $\br^t$ in the transmit surface $S$}, then the radiated electric field $\bE(\br^r)$ at the location $\br^r$ in the half free-space is given by the dyadic {Green’s function theorem as \cite{Wei2023TWC},{\cite{Mikki2011TAP}}}
\begin{equation}
{\bE(\br^r)=i\omega\mu\int_{S} \bG(\br^t,\br^r) \bJ(\br^t) \text{d}s }, \label{eq:Green1}
\end{equation}
{where $\omega$ is permittivity, $\mu$ is permeability, and {$\bG(\br^t,\br^r)$} is dyadic Green’s function} \cite{Wei2023TWC},\cite{Guo2024Trans}, {\cite{arnoldus2001representation}}, {\cite{Mikki2018Access}}
\begin{eqnarray}
\bG(\br^t,\br^r)= g(\br^t,\br^r)  \left[c_1(r)\bI_3+c_2(r) \vec\br \vec\br\tra\right], \label{eq:bG} 
\end{eqnarray}
where 
\begin{eqnarray}
c_1(r)=1+\frac{i}{k_0 r}-\frac{1}{k_0^2 r^2}, \ \  
c_2(r)=\frac{3}{k_0^2 r^2}-\frac{3i}{k_0 r}-1,  \nonumber
\end{eqnarray}
the unit vector $\vec\br={{(\br^t-\br^r)}}/{||\br^t-\br^r||}$ denotes the direction between the source point
and observation point, the scalar $r=||\br^t-\br^r||$ denotes their distance, $\bI_{3}$ denotes an identity matrix, $k_0=2\pi/\lambda$ is the wave number with $\lambda$ being the wavelength, 
and the scalar Green’s function $g(\br^t,\br^r)$ is given as 
\begin{eqnarray}
g(\br^t,\br^r)=\frac{\exp(ik_0||\br^t-\br^r||)}{4\pi ||\br^t-\br^r||}. \label{eq:scalarG}
\end{eqnarray}

Now we consider a single transmit patch $S_n^t$ and a single receive patch $S_m^r$, and assume that the current distribution over the transmit patch is {constant \cite{Wei2023TWC}}. 
Then the wireless channel with polarization between the $n$-th transmit patch and the $m$-th receive patch can be expressed as
\begin{eqnarray}
\bar\bH_{mn} &=&i\omega\mu 
\int_{x^r_m-\frac{\Delta_x^r}{2}}^{x^r_m+\frac{\Delta_x^r}{2}}
{\int_{y^r_m-\frac{\Delta_y^r}{2}}^{y^r_m+\frac{\Delta_y^r}{2}} 
\int_{x^t_n-\frac{\Delta_x^t}{2}}^{x^t_n+\frac{\Delta_x^t}{2}} }
\int_{y^t_n-\frac{\Delta_y^t}{2}}^{y^t_n+\frac{\Delta_y^t}{2}} \nonumber\\
&&\ \ \  \bG(\br^t,\br^r) \text{d}y^t\text{d}x^t\text{d}y^r\text{d}x^r. \label{eq:Green22}
\end{eqnarray}
We note that $\bar\bH_{mn}$ is a matrix with size $3 \times 3$, i.e., 
\begin{eqnarray}
\bar\bH_{mn} =
\begin{bmatrix}
h_{mn}^{xx} & h_{mn}^{xy} & h_{mn}^{xz} \\
h_{mn}^{xy} & h_{mn}^{yy} & h_{mn}^{yz} \\
h_{mn}^{xz} & h_{mn}^{yz} & h_{mn}^{zz} 
\end{bmatrix},\label{eq:Mtx_Hmn}
\end{eqnarray}
where $h_{mn}^{\kappa}, {\kappa\in \{xx,xy,xz, yy, yz, zz \}}$ denotes the channel component corresponding to various transmit and receive polarization combinations. 
Considering all transmit and receive patches, we have the following channel matrix
\begin{eqnarray}
\tilde\bH \triangleq
\begin{bmatrix}
\bar\bH_{11} & \cdots &\bar\bH_{1N}  \\
\vdots   &\ddots  &\vdots \\
\bar\bH_{M1} &\cdots  &\bar\bH_{MN} 
\end{bmatrix}
\in \mathbb{C}^{3M\times 3N}. 
\end{eqnarray}
We can rearrange the elements in the matrix according to their polarization combinations, leading to a block matrix 
\begin{eqnarray}
\tilde\bH =
\begin{bmatrix}
\tilde\bH_{xx} & \tilde\bH_{xy} &\tilde\bH_{xz}  \\
\tilde\bH_{xy} &\tilde\bH_{yy} &\tilde\bH_{yz} \\
\tilde\bH_{xz} &\tilde\bH_{yz} &\tilde\bH_{zz} 
\end{bmatrix}. \nonumber
\end{eqnarray}
where $\tilde\bH_{\kappa}\in \mathbb{C}^{M\times N}$ denotes the polarized channel matrix with $\kappa \in \{xx,xy,xz,yy,yz,zz\}$. 


\subsection{{Problem Formulation for Channel Estimation}}

{Arranging $L$ consecutive received signal vectors in a matrix $\tilde\bY$, we have 
\begin{eqnarray}
\tilde\bY=\tilde\bH\tilde\bS+\tilde \bW, \label{eq:Recv0} 
\end{eqnarray}
where $\tilde\bY \triangleq [\tilde\bY_x^T, \tilde\bY_y^T, \tilde\bY_z^T]^T \in \mathbb{C}^{3M\times L}$ with the subscripts denoting the polarization direction and $\tilde\bY_x, \tilde\bY_y, \tilde\bY_z \in \mathbb{C}^{M\times L}$.
$\tilde\bS \triangleq [\tilde\bS_x^T, \tilde\bS_y^T, \tilde\bS_z^T]^T \in \mathbb{C}^{3N\times L}$ with $\tilde\bS_x, \tilde\bS_y, \tilde\bS_z \in \mathbb{C}^{N\times L}$ denoting the pilot matrices in $x,y$ and $z$ polarization, 
$\tilde \bW \triangleq[\tilde\bW_x^T,\tilde\bW_y^T, \tilde\bW_z^T]^T\in \mathbb{C}^{3M\times L}$ represents the zero mean complex additive white Gaussian noise (AWGN) with precision $\gamma$ (i.e., variance $\gamma^{-1}$). 


To facilitate channel estimation, we can rearrange the signal model into the following form
\begin{eqnarray}
\bY&&=
\begin{bmatrix}
\tilde\bS_x\tra & 0 & 0  & \tilde\bS_y\tra  & \tilde\bS_z\tra  & 0  \\
0 & \tilde\bS_y\tra & 0  & \tilde\bS_x\tra  & 0  & \tilde\bS_z\tra  \\
0 & 0 & \tilde\bS_z\tra  & 0  & \tilde\bS_x\tra  & \tilde\bS_y\tra 
\end{bmatrix}
\begin{bmatrix}
\bH_{xx} \\ \bH_{yy} \\ \bH_{zz} \\ \bH_{xy} \\ \bH_{xz} \\ \bH_{yz} 
\end{bmatrix}
+\bar{\bW} \nonumber\\ 
&&\triangleq {\bS} \bH+\bar{\bW},\label{eq:Recv2}
\end{eqnarray}
{where $\bY\triangleq[\tilde\bY_x,\tilde\bY_y,\tilde\bY_z]^T{\in\mathbb{C}^{3L\times M}}$, $\bS\in \mathbb{C}^{3L\times 6N}$, $\bH\in \mathbb{C}^{6N\times M}$ 
and $\bH_{\kappa}=\tilde\bH_{\kappa}^T$ with $\kappa \in \{xx,xy,xz,yy,yz,zz\}$, which is given as
\begin{eqnarray}\label{eq:Hkappa}
\bH_{\kappa} \triangleq
\begin{bmatrix}
h^{\kappa}_{11} & \cdots & h^{\kappa}_{1M} \\
\vdots          & \ddots & \vdots \\
h^{\kappa}_{N1} &  \cdots & h^{\kappa}_{NM} \\
\end{bmatrix}.
\end{eqnarray} 
In addition, $\bar{\bW}\triangleq[\tilde\bW_x, \tilde\bW_y, \tilde\bW_z]^T\in \mathbb{C}^{3L\times M}$ denotes the white Gaussian noise. 

Our aim is to estimate the channel matrix $\bH$ based on the pilot signals $\bS$ and received signal $\bY$. Regarding this, we have the following remarks:  
\begin{itemize}
\item One straightforward method is to estimate the channel coefficients directly e.g., using the least squares (LS) method. It is noted the number of the variables to be estimated is $6MN$, which lead to high pilot overhead to achieve satisfactory performance and high computational complexity due to the involved large matrix inversion. 
\item According to \eqref{eq:Green22}, the channel coefficients are parameterized by $\br^r$ and $\br^t$ (noting that $\br^r$ is known). This motivates us to perform parametric channel estimation to drastically reduce the number of variables to be estimated, thereby achieving significantly enhanced performance. That is, we first estimate $\br^t$, based on which the channel matrix can be reconstructed. This is the strategy of parametric channel estimation used in this work.  
\item It can be seen from \eqref{eq:Green22} that 
there is a complex relationship between {$\bar\bH_{mn}$} and $\br^t$, leading to challenges in parametric channel estimation.  
In \cite{Wei2023TWC}, with some approximations, an approximate analytical expression for the HMIMO channel is obtained, i.e.,  
\begin{eqnarray}
&&\!\!\!\!\!\!\!\!\!\!\!\!\!\!\!\bar\bH_{mn} \approx i\omega\mu\Delta^t \Delta^r \frac{\exp(ik_0r_{mn})}{4\pi r_{mn}}\times \nonumber\\
&&\!\!\!\!\!\!\!\!\!\ \text{sinc}\frac{k_0(x_m^r-x_n^t)\Delta_x^t}{2r_{mn}}\text{sinc}\frac{k_0(y_m^r-y_n^t)\Delta_y^t}{2r_{mn}}\bC_{mn}, \label{Appmod}
\end{eqnarray} 
{where $\Delta^t=\Delta^t_x\Delta^t_y$ and $\Delta^r=\Delta^r_x\Delta^r_y$ represent the area of transmitter and receiver patch, respectively, $r_{mn}=||\br^t_n-\br^r_m||$ and $\bC_{mn}=c_1(r_{mn})\bI_3+c_2(r_{mn}) \vec\br \vec\br^T$ with $c_1(r_{mn})$ and $c_2(r_{mn})$ are given in \eqref{eq:bG} by replacing $\br^t$ and $\br^r$ with the $\br^t_n$ and $\br^r_m$, respectively.}  However, the approximate channel model still has a complex expression with nonlinear operations, making parametric channel estimation challenging. Moreover, it can also lead to significant mismatch with the true channel due to the approximations. 
\item In this work, we propose a novel NN-assisted hybrid channel model to characterize the nonlinear relationship between $\bH$ and $\br^t$, which has a much simpler expression, enabling efficient parametric channel estimation.
\end{itemize}



\begin{figure}[!t]
\centering
\includegraphics[width=0.5\textwidth]{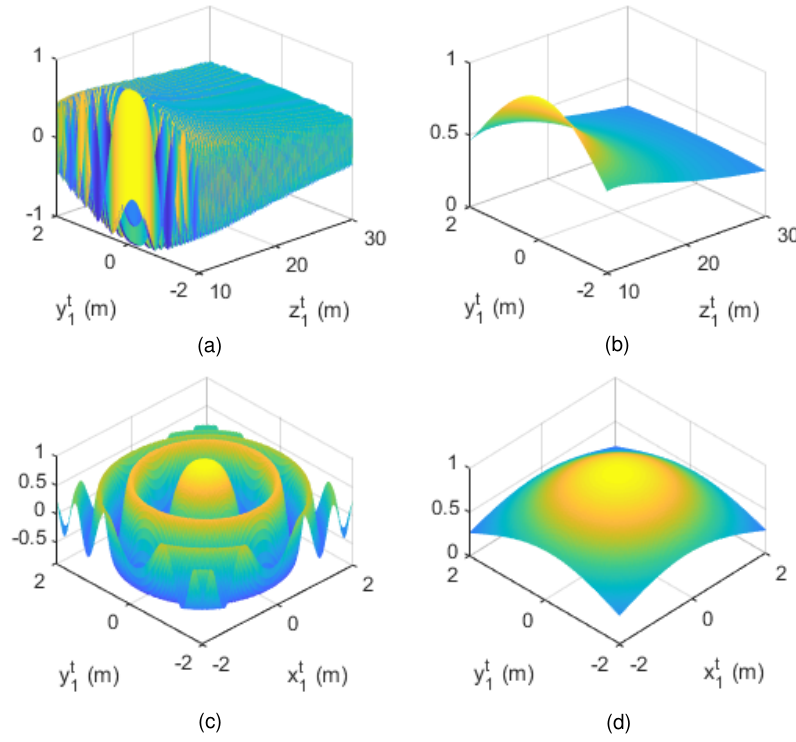}\centering
\caption{The real part of $h^{xx}_{11}$ (a) and (c) and the real part of  $h^{xx}_{11}/\exp(ik_0 r_{11})$ (b) and (d).}   
\label{fig:ChannelHxx}
\end{figure}

\section{ NN-Assisted Hybrid Channel Model and Parametric Channel Estimation}
\subsection{NN-Assisted Hybrid Channel Model}
According to \eqref{eq:Green22}, the channel components between the $m$-th receive patch and the $n$-th patch $h^{\kappa}_{mn}$ is parameterized by $\{x^t_n,y^t_n,z^t_n\}$ (noting that the coordinates of the receive patch centers are known). An idea is to develop an NN model with the parameters as input modes to replace \eqref{eq:Green22}. However, this is not the best way. 
Take the channel component $h^{\kappa}_{11}$, which corresponds to the 1st transmit patch and the 1st receive patch, as an example.  In Fig. \ref{fig:ChannelHxx} (a) and (c), we show the real part of $h^{xx}_{11}$ respectively with a fixed $x_1^t$ and a fixed $z_1^t$. 
We can see that they are rather complex and change abruptly, which makes it necessary to use NNs with high expressive capability, leading to the requirement of a large number of NN parameters. This will result in high complexity in training and parametric channel estimation channel estimation.

The decayed periodicity-like pattern exhibited in Fig. \ref{fig:ChannelHxx} (a) and (c) motivates us to examine the variation of the quantity 
\begin{equation}
\tilde{h}^{xx}_{11}=h^{xx}_{11}(x_1^t, y_1^t, z_1^t)/\exp(ik_0 r_{11}),
\end{equation}
where $r_{11}=||\br_1^t-\br_1^r||$.  
Its real part is shown in Fig. \ref{fig:ChannelHxx} (b) and (d), where we can see that it changes much slowly. Hence, it can be potentially characterized by a much simpler NN. In particular, we use an NN with a single hidden layer as shown in Fig. \ref{fig:NeuralNetwork} in this work. 

It is not hard to show that the channel components between the $m$th receive patch and the $n$th transmit patch actually depends on their relative position.
Then we define
${x_{mn}}=x^{t}_{n}-x^{r}_{m}, {y_{mn}}=y^{t}_{n}-y^{r}_{m}, {z_{mn}}=z^{t}_{n}-z^{r}_{m}$
$\br_{mn}=[x_{nm},y_{nm},z_{nm}]^T$ and $r_{mn}=||\br_{mn}||$. Hence, we have
\begin{equation} 
\tilde{h}^{\kappa}_{mn}=h^{\kappa}_{mn}(x_{mn}, y_{mn}, z_{mn})/\exp(ik_0 r_{mn}),  
\label{eq:neweq} 
\end{equation}
where $\kappa \in \{xx,xy,xz,yy,yz,zz\}$. This allows us to use a single NN to characterize the channel components between any transmit patch and receive patch, facilitating the NN training and Bayesian inference algorithm design later.   

As shown in Fig. \ref{fig:NeuralNetwork}, we use a real-valued NN, where we separate the real and imaginary parts of the relevant variables. Its inputs are the relative coordinates $x_{mn}, y_{mn}$ and $z_{mn}$, and the outputs are $\{Re\{\tilde{h}^{\kappa}_{mn}\}, Im\{\tilde{h}^{\kappa}_{mn}\}, \kappa \in \{xx,xy,xz,yy,yz,zz\}\}$.  According to Fig. \ref{fig:NeuralNetwork}, the output of the neural network can be expressed as
\begin{align}
&\mathcal{NN}(x_{mn}, y_{mn}, z_{mn})\nonumber\\
&=\bW_2\tra g_a(\bw_1^x x_{mn}+\bw_1^y y_{mn}+\bw_1^z z_{mn}+\bb_1)+\bb_2, 
\label{eq:neuralnetwork}
\end{align}
where $\bW_1 = [\bw_1^x,\bw_1^y,\bw_1^z]\in \mathbb{R}^{N_h\times 3}$ is the input layer weight matrix with $\bw_1^x,\bw_1^y,\bw_1^z\in \mathbb{R}^{N_h\times 1}$, $\bW_2 = [\bw_{2,1},\cdots,\bw_{2,12}] \in\mathbb{R}^{N_h\times 12}$ is the output layer weight matrix  with $\{\bw_{2,1},\cdots,\bw_{2,12}\}\in \mathbb{R}^{N_h\times 1}$,  output  $\mathcal{NN}(x_{mn}, y_{mn}, z_{mn})\in\mathbb{R}^{12\times 1}$,
$N_h$ is the number of neurons in the hidden layer, $\bb_1 \in \mathbb{R}^{N_h\times 1}$ and $\bb_2\in \mathbb{R}^{12\times 1}$ are bias vectors in the hidden layer and output layer respectively. The activation function in the hidden layer $g_a(\cdot)=tansig(\cdot)$ and a linear activation function is used at the output layer. 

\begin{figure}[!t]
\centering
\includegraphics[width=0.45\textwidth]{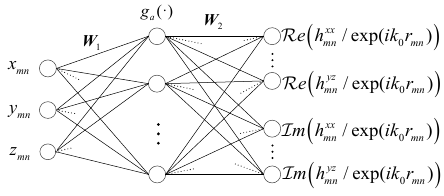}
\centering
\caption{Architecture of the neural network.} 
\label{fig:NeuralNetwork}
\end{figure}

We put two outputs in Fig. \ref{fig:NeuralNetwork}, which correspond to the real and imaginary parts of $\tilde{h}^{\kappa}_{mn}$ together and denote it as ${\varphi}^{\kappa}(x_{mn},y_{mn},z_{mn})$. Then we have the following hybrid channel model
\begin{eqnarray}
h_{mn}^{\kappa} 
\approx {\varphi}^{\kappa}(x_{mn},y_{mn},z_{mn}) \exp(ik_0 r_{mn}), \label{eq:Hmn_NN}
\end{eqnarray}
where ${\varphi}^{\kappa}(x_{mn},y_{mn},z_{mn})$ is the output of the NN. We can see that, with the aid of NN, we convert the channel model with complex expression {\eqref{eq:Green22}} to {\eqref{eq:Hmn_NN}}, which has a closed form with much simpler expression. It is noted that all the operations involved in {\eqref{eq:neuralnetwork}} and {\eqref{eq:Hmn_NN}} are linear, except the nonlinear operations due to the activation function {$g_a(\cdot$)} in the NN and multiplication operation in {\eqref{eq:Hmn_NN}}.  
The hybrid channel model enables tractable Bayesian inference for parametric channel estimation detailed in Section IV.


According to \eqref{eq:coord_xmn}, the relative coordinates can be expressed as functions of $\{x_1^t, y_1^t, z_1^t\}$ or $\{x_n^t, y_n^t, z_n^t\},\forall n$, i.e., 
\begin{eqnarray}
x_{mn}&=&x_n^t-x_m^r=x_1^t+(c_n^t-1)\Delta_x^t-x_m^r{=x_1^t+\Delta_{mn}^x}  \nonumber  \\
y_{mn}&=&y_n^t-y_m^r=y_1^t+(l_n^t-1)\Delta_y^t-y_m^r{=y_1^t+\Delta_{mn}^y}  \nonumber \\
z_{mn}&=&z_n^t=z_1^t. 
\label{eq:coordinate}
\end{eqnarray}
It is noted that the coordinates of the receive patches $\{x_m^r, y_m^r \}$ are known. This means that any channel component $h_{nm}^{\kappa}$ can be expressed as a function of the coordinate of the first transmit patch $\{x_1^t, y_1^t, z_1^t\}$, which will be estimated. With the estimated coordinate, the channel components can be obtained using \eqref{eq:Hmn_NN}. 
We define a new function ${\phi}^{\kappa}_{mn}(x_{n}^t,y_{n}^t,z_{n}^t)$, which is a shifted version of ${\varphi}^{\kappa}(x_{mn},y_{mn},z_{mn})$, i.e.,
\begin{eqnarray} 
{\phi}^{\kappa}_{mn}(x_{n}^t,y_{n}^t,z_{n}^t)={\varphi}^{\kappa}(x_{n}^t-x_m^r,y_{n}^t-y_m^r,z_{n}^t), \label{eq:Hmn_NN2}
\end{eqnarray}
which will be used later in the inference algorithm design.

\begin{figure}[!t]
\centering
\includegraphics[width=0.4\textwidth]{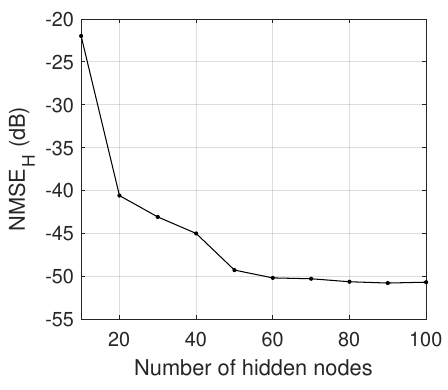}
\centering
\caption{NMSE versus the number of hidden nodes.}
\label{fig:MSEvsNh}
\end{figure}    

\subsection{{Neural Network Training}}
The training of the NN can be easily implemented. This is because, with the original model \eqref{eq:Green22} 
sufficient training samples can be easily obtained using numerical methods. We note that the whole training process can be carried out offline. 
The NN is trained with the back propagation using the following loss function
\begin{equation}
\text{L}= \frac{1}{MN}\sum_{m,n, \kappa} ||{\varphi}^{\kappa}_{mn}(x_{mn},y_{mn},z_{mn})-\tilde{h}^{\kappa}(x_{mn},y_{mn},z_{mn})||^2\nonumber
\end{equation} 
where 
$\tilde{h}^{\kappa}(x_{mn},y_{mn},z_{mn})$ are calculated using \eqref{eq:Green22} and \eqref{eq:neweq}.

The number of hidden nodes impacts the accuracy of the NN model. Fig. \ref{fig:MSEvsNh} show the normalized mean squared error (NMSE) performance of the predicted channel versus the number of hidden nodes. We can see that the NMSE performance improves with the number of hidden nodes. When the number of the hidden nodes is larger than 50, the NMSE reaches about -50dB and the performance improvement is marginal with the further increase of the hidden node number. 
Hence, we choose the number of hidden nodes to be 50.

\section{Probabilistic Formulation, Factor Graph Representation and Message Passing Algorithm}\label{Sec:MP}

\subsection{Probabilistic Formulation} 
We develop a Bayesian parametric channel estimation method leveraging UAMP to achieve low complexity while with high robustness. 
To facilitate the use of UAMP, we carry out a unitary transformation to \eqref{eq:Recv2} based on the singular value decomposition (SVD) $\bS=\bU\bLambda\bV$, i.e., 
\begin{eqnarray}
\bR=\bPhi\bH+\bW, \label{eq:Recv3}
\end{eqnarray}
where $\bR=\bU^H \bY$, $\bPhi=\bU^H \bS$, and $\bW=\bU^H \bar{\bW}$ is still zero-mean white Gaussian noise with the same variance because $\bU^H$ is a unitary matrix.


The conditional joint probability density function of the unknown variables given the observation matrix $\bR$ can be factorized as
\begin{align}
&p(\bH, \gamma, {\{x^t_{n}, y^t_{n}, z^t_{n}, \forall n\}, x_1^t, y_1^t, z_1^t} |\bR)  \nonumber\\
&\propto p(\bR|\bH,\gamma)\prod\nolimits_{n,m} p(\bh_{mn}|x^t_{n}, y^t_{n}, z^t_{n})p(x^t_{n}|x_1^t) \nonumber\\
&\ \ \ \ \ \times p(y_{n}|y_1^t) p(z_{n}|z_1^t) p(x_1^t) p(y_1^t) p(z_1^t) p(\gamma) \nonumber\\
&\ \triangleq f_{\bR}(\bR,\bH,\gamma)
\prod\nolimits_{n,m} f_{h_{mn}}(\bh_{mn},x^t_{n}, y^t_{n}, z^t_{n})f_{x^t_n}(x^t_n,x^t_1) \nonumber\\
&\times f_{y^t_n}(y^t_n,y^t_1)f_{z^t_n}(z^t_n,z^t_1)f_{x^t_1}(x_1^t)f_{y^t_1}(y_1^t)f_{z^t_1}(z_1^t)f_\gamma(\gamma), \label{eq:factorization}
\end{align}
where $\gamma$ is the precision of the noise and it is treated as a random variable with an improper prior $p(\gamma)\propto 1/\gamma$, the function $f_{\bR}(\bR,\bH,\gamma) 
=\CN(\bR;\boldsymbol{\Phi}\bH,\gamma^{-1}\bI)$, 
the function $f_{h_{mn}}(\bh_{mn},x_n^t,y_n^t,z_n^t)$ can be further decomposed into
\begin{align}
f_{h_{mn}}(\bh_{mn},x_n^t,y_n^t,z_n^t)=\prod_\kappa f_{ h_{mn}}^\kappa(h_{mn}^\kappa,x_n^t,y_n^t,z_n^t), \nonumber
\end{align}
with {$\bh_{mn}= [h_{mn}^{xx},...,h_{mn}^{zz}]^T\in\mathbb{C}^{6\times 1}$} and $f_{ h_{mn}}^\kappa (x_n^t,y_n^t,z_n^t)= \delta(h_{mn}^\kappa-\phi_{mn}^\kappa  \exp(ik_0r_{mn}))$ (the arguments $x_n^t, y_n^t, z_n^t$ of the function $\phi_{mn}^\kappa$ are omitted for notation simplicity), the function $f_{x^t_n}(x^t_n,x^t_1)=\delta\left(x_n^t-(x_1^t+(c_n^t-1)\Delta_x^t) \right)$ according to \eqref{eq:coordinate}, 
{$f_{x_1}(x_1^t)$ represents the prior of $x_1^t$, which is selected to be a non-informative one, e.g., a Gaussian distribution with an infinite variance.} Other functions $f_{y^t_n}(y^t_n,y^t_1)$, $f_{z^t_n}(z^t_n,z^t_1)$, $f_{y^t_1}(y_1^t)$ and $f_{z^t_1}(z_1^t)$ have similar definitions as $f_{x^t_n}$ and $f_{x_1}$, as shown in Table I. Our aim is to compute the (approximate) marginals of the coordinates $x^t_1, y^t_1, z^t_1$ and the channel components, based on which their estimates can be obtained.

\begin{table}[hbp]
\centering
\renewcommand\arraystretch{1.2}
\caption{Local functions and distributions in \eqref{eq:factorization} }
\begin{tabular}{ p{35pt} p{65pt} p{125pt}}
\hline
Factor & Distribution & Function\\
\hline
$f_{\bR} $ & $p(\bR|\bH,\gamma)$ & $\CN(\bR;\boldsymbol{\Phi}\bH,\gamma^{-1}\bI)$\\
$f_{h_{mn}}^\kappa $ & $p(\bh_{mn}|x_n^t,y_n^t,z_n^t)$ & {$\prod_\kappa \delta( h_{mn}^\kappa-\phi_{mn}^\kappa \exp(ik_0r_{mn}))$}\\
$f_{x^t_n}$ & $p(x_{n}|x_1^t)$ & $\delta\left(x_n^t-(x_1^t+(c_n^t-1)\Delta_x^t) \right)$ \\
$f_{y^t_n}$ & $p(y_{n}|y_1^t)$ & $\delta\left(y_n^t-(y_1^t+(l_n^t-1)\Delta_y^t) \right)$ \\
$f_{z^t_n}$ & $p(z_{n}|z_1^t)$ & $\delta\left(z_n^t-z_1^t \right)$ \\
{$f_{x^t_1}$,$f_{y^t_1}$,$f_{z^t_1}$}& $p(x_1^t),p(y_1^t),p(z_1^t)$ & non-informative \\
$f_{\gamma}$ & $p(\gamma)$ & $1/\gamma$ \\
\hline
\label{Table}
\end{tabular}
\end{table}

\begin{figure}[!t]
\centering
\includegraphics[width=0.45\textwidth]{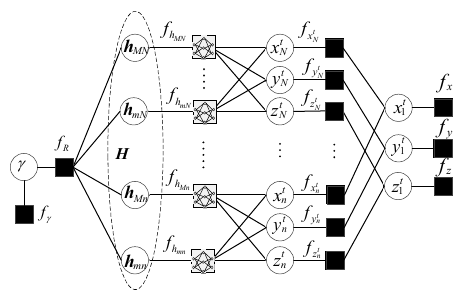}
\centering
\caption{Factor graph representation of  \eqref{eq:factorization}}. 
\label{fig:FactorGraph}
\end{figure}

\subsection{{Factor Graph Representation}}

To facilitate the factor graph representation of the factorization in \eqref{eq:factorization}, we list the involved notations in Table \ref{Table}, showing the correspondence between the factor labels and the underlying distributions they represent, and the specific functional form assumed by each factor. In this section, we investigate how to efficiently solve the formulated channel estimation problem with message passing-based Bayesian inference. 

The factor graph representation for the factorization in \eqref{eq:factorization} is visualized in Fig. \ref{fig:FactorGraph}, where squares and circles represent function nodes and variable nodes, respectively. It is noted that the message passing algorithm is performed in an iterative manner, where each iteration involves a forward message passing process and backward message passing process in the graph shown in Fig. \ref{fig:FactorGraph}. We use $m_{A\rightarrow B}(\mu)$ to denote a message passed from node $A$ to node $B$, which is a function of $\mu$. For Gaussian messages, the arrows above its mean and variance indicate the message passing direction. In addition, we use $b(\mu)$ to denote the belief of a variable $\mu$. Note that, if a forward computation requires backward messages, the relevant messages in the previous iteration is used by default.
Next we elaborate the forward and backward message computations. {To facilitate derivations, a scalar representation of the part of the graph is shown in Fig. \ref{fig:scalarNNgraph} to show the detailed relationship between the hybrid function nodes and variable nodes.}

\begin{figure}[!t]
\centering
\includegraphics[width=0.3\textwidth]{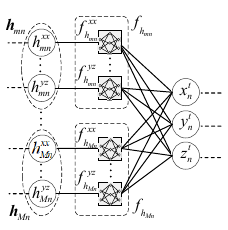}
\centering
\caption{Scalar factor graph representation related to the hybrid local function nodes.}
\label{fig:scalarNNgraph}
\end{figure}

\subsection{{ Forward Message Passing}}
{From the factorization \eqref{eq:factorization}, the likelihood function $p(\bR|\bH,\gamma)$ is a linear mixed model and can be handled using UAMP \cite{guo2015approximate, yuan2020approximate,  Luo2021Tsp}. With UAMP, we can compute the mean and variances about the entries in matrix $\bH$ in the following. 
Following UAMP, we first compute matrices $\bV_{P}$ and $\bP$ as} 
\begin{eqnarray}
\bV_{P}&=&|\bPhi|^{.2} \bV_H, \nonumber\\
\bP&=&\bPhi\hat\bH-\bV_{P}\cdot \bS_H, \nonumber
\end{eqnarray}
where $\hat\bH$ is the mean matrix of $\bH$ and $\bV_H$ contains the variances of the corresponding elements in $\bH$, which will be updated in \eqref{eq:blf_h} based on the posterior distribution $b(\bH)$, and $\bS_H$ will be updated \eqref{eq:SH}.
The precision of the noise $\gamma$ can be estimated as
\begin{eqnarray}
\hat\gamma=\frac{3ML}{||\bR-\bZ||^2+\bV_{Z}}, \label{eq:gamma}
\end{eqnarray}
where the auxiliary matrices $\bV_{Z}$ and $\bZ$ can be computed as
\begin{align}
\bV_{Z}&=\bV_{P}./(\hat\gamma \bV_{P}+1), \\
\bZ_E&=(\hat\gamma \bR+\bP./\bV_{P})\cdot\bV_{Z}.
\end{align}
Then, we update intermediate matrices $\bV_{S_H}$ and $\bS_H$ as
\begin{eqnarray}
&&\bV_{S_H}=1./(\bV_{P}+\hat\gamma^{-1}), \nonumber\\ 
&&\bS_H=\bV_{S_H}\cdot (\bR-\bP),\label{S_H}
\end{eqnarray}
and obtain matrices $\bV_{Q_H}$ and $\bQ_H$ as
\begin{align}
\bV_{Q_H}=&1./(|\bPhi^H|^{.2}\bV_{S_H}),\\
\bQ_H=&\hat\bH+\bV_{Q_H}\cdot(\bPhi^H\bS_H).\label{eq:Q}
\end{align}
Matrices $\bQ_H$ and $\bV_{Q_H}$ can be respectively represented as
\begin{eqnarray}
\bQ_H=
\begin{bmatrix}
\bQ_{H_{xx}}\\
\vdots \\
\bQ_{H_{yz}}
\end{bmatrix},
\ \ \ 
\bV_{Q_H}=
\begin{bmatrix}
\bV_{Q_{H}^{xx}}\\
\vdots \\
\bV_{Q_{H}^{yz}}
\end{bmatrix}.
\end{eqnarray}
where $\bQ_{H_{\kappa}}\in\mathbb{C}^{N\times M}$ and $\bV_{Q_H^\kappa}\in\mathbb{R}^{N\times M}, \kappa\in\{xx,...,zz\}$ with their $(n,m)$-th element denoted as $q_{mn}^{\kappa}$ and $\nu_{q_{mn}^{\kappa}}$, respectively. Here $q_{mn}^{\kappa}$ and $\nu_{q_{mn}^{\kappa}}, \forall \kappa$, represent the mean and variance of message $m_{h^{\kappa}_{mn}\to f_{h^{\kappa}_{mn}}}(h^{\kappa}_{mn})$, i.e., 
\begin{eqnarray}\label{eq:hknm2fhknm}
m_{h^{\kappa}_{mn}\to f_{h^{\kappa}_{mn}}}(h^{\kappa}_{mn})=\CN(h^{\kappa}_{mn}; q^{\kappa}_{mn}, \nu_{q^{\kappa}_{mn}}).\label{eq:msg_q}
\end{eqnarray}

From the factor graph, we can see that the message updates for {$x_1^t, y_1^t$ and $z_1^t$} are similar. So in the following, we take {$x_1^t$} as an example for the derivation of the message update rules.
{It can be seen from \eqref{eq:Hmn_NN} that the local function node $f_{h_{mn}}^\kappa$ still involves nonlinear operations, leading to intractable messages.  
To overcome the problem, we propose using Taylor expansion to dynamically linearize the node $f_{h_{mn}}^\kappa$. To this end, we approximate $h_{mn}^{\kappa}$ as 
\begin{align}
&h_{mn}^{\kappa}(x_{n}^t,y_{n}^t,z_{n}^t) \nonumber\\
&\approx \underbrace{h_{mn}^{\kappa}(\hat x_{n}^t,\hat y_{n}^t,\hat z_{n}^t)}_{\hat h_{mn}^{\kappa}}+
\underbrace{h_{mn}^{\kappa,x'}(\hat x_{n}^t,\hat y_{n}^t,\hat z_{n}^t)}_{\hat h_{mn}^{\kappa,x'}} (x_{n}^t-\hat x_{n}^t)\nonumber\\
&\ \ \ \ +\underbrace{h_{mn}^{\kappa,y'}(\hat x_n^t,\hat y_n^t,\hat z_n^t)}_{\hat h_{mn}^{\kappa,y'}} (y_n^t-\hat y_n^t)+
\underbrace{h_{mn}^{\kappa,y'}(\hat x_n^t,\hat y_n^t,\hat z_n^t)}_{\hat h_{mn}^{\kappa,z'}}(z_n^t-\hat z_n^t)\nonumber\\
& =\underbrace{\hat h_{mn}^{\kappa}-\hat h^{\kappa, x'}_{mn}\hat x^t_n-\hat h^{\kappa, y'}_{mn}\hat y_n^t-\hat h^{\kappa, z'}_{mn}\hat z_n^t}_{\xi_{mn}^{\kappa}}\nonumber\\
&\ \ \ +x_n^t\hat h^{\kappa, x'}_{mn}+y_n^t\hat h^{\kappa, y'}_{mn}+z_n^t\hat h^{\kappa, z'}_{mn}\nonumber\\
&=\xi_{mn}^{\kappa}+x_n^t\hat h^{\kappa, x'}_{mn}+y_n^t\hat h^{\kappa, y'}_{mn}+z_n^t\hat h^{\kappa, z'}_{mn}. \label{eq:nn_approx}
\end{align}
where $h_{mn}^{\kappa,x'}$ represents the partial derivative of $h_{mn}^{\kappa}$ with respect to $x_n^t$, and similarly, $h_{mn}^{\kappa,y'}$ and $h_{mn}^{\kappa,z'}$ are the partial derivatives with respect to $y_n^t$ and $z_n^t$, respectively. 
}
{The partial derivative of $h_{mn}^{\kappa,x'}$ can be obtained as 
\begin{eqnarray}
h_{mn}^{\kappa,x'}=\frac{\partial h_{mn}^{\kappa}}{\partial x_n^t}=\left( \frac{\partial\phi^{\kappa}_{mn}}{\partial x_n^t} +\phi_{mn}^{\kappa}ik_0 \frac{\partial r_{mn}}{\partial x_n^t}\right) {\exp(ik_0 r_{mn})}.
\end{eqnarray}
The derivatives of $\phi_{mn}^{\kappa}$ can be get by
\begin{eqnarray}
&&\frac{\partial \phi^{\kappa}_{mn}}{\partial x_n^t }
=\left((\bw_{2,\kappa_1}+\bw_{2,\kappa_2})\cdot \bw_1^x\right)\tra \nonumber\\ 
&&\ \ \ \ \ \ \ \ \ \ \ \ \times g'_a(x_n^t\bw_1^x+y_n^t\bw_1^y+z_n^t\bw_1^z+\bb_1),\nonumber
\end{eqnarray}
{where the values of indices $\kappa_1$ and $\kappa_2$ depend on the polarization parameter $\kappa \in \{xx,...,zz\}$}. For example, if $\kappa=xx$, we have $\kappa_1=1$ and $\kappa_2=7$. The derivative of $g_a(\cdot)$ is $g'_a(\cdot)=1-g^2_a(\cdot)$.}

{According to \eqref{eq:nn_approx}, we can rewrite the function $f_{h_{mn}^\kappa}$ as
\begin{eqnarray}
f_{h_{mn}^\kappa}=\delta\left(h_{mn}^\kappa-(\xi_{mn}^{\kappa}+x_n^t\hat h^{\kappa, x'}_{mn}+y_n^t\hat h^{\kappa, y'}_{mn}+z_n^t\hat h^{\kappa, z'}_{mn})\right).
\end{eqnarray}
Assume that the backward messages $m_{y^t_n \to f_{h_{mn}}^\kappa}(y^t_n)$ and $m_{z^t_n\to f_{h_{mn}}^\kappa}(z^t_n)$ are available, which turn out to be Gaussian (please refer to a similar message $m_{x^t_n\to f_{h_{mn}}^\kappa}(x^t_n)$ in \eqref{eq:msg_u2fh} that is derived later) and can be expressed as
\begin{align}
m_{y^t_n \to f_{h_{mn}}^\kappa}(y^t_n)=\CN\left(y^t_n; \cev y^\kappa_{mn}, \cev\nu_{y^\kappa_{mn}}\right),\\
m_{z^t_n\to f_{h_{mn}}^\kappa}(z^t_n)=\CN\left(z^t_n; \cev z^\kappa_{mn} \cev\nu_{z^\kappa_{mn}}\right).\label{eq:yzBack}
\end{align}
Then the message from $f_{h_{mn}^\kappa}$ to $x^t_n$ can be computed as
\begin{align}
m_{f_{h_{mn}^\kappa}\to x^t_n}(x^t_n)&=\int f_{h_{mn}^\kappa} m_{y^t_n\to f_{h_{mn}}^\kappa}(y^t_n) m_{z^t_n\to f_{h_{mn}}^\kappa}(z^t_n) \nonumber\\
&\ \ \ \ \ \ \times m_{h_{mn}^\kappa\to f_{h_{mn}^\kappa}}(h_{mn}^\kappa)\text{d} y^t_n\text{d} z^t_n\text{d}h_{mn}^\kappa \nonumber\\
&=\CN\left(x^t_n; \vec x_{mn}^\kappa, \vec\nu^{\kappa}_{{x_{mn}}}\right),\label{eq:msg_h2u}
\end{align}
where
\begin{eqnarray}
\vec x_{mn}^\kappa=\frac{q_{mn}^\kappa-\xi^\kappa_{mn}-\cev y_{mn}^\kappa \hat h^{\kappa y'}_{mn}-\cev z_{mn}^\kappa\hat h^{\kappa z'}_{mn}} {\hat h^{\kappa x'}_{mn}},\\
\vec\nu^{\kappa}_{x_{mn}}=\frac{\nu_{q_{mn}^\kappa}+\cev\nu_{y_{mn}^\kappa}|\hat h^{\kappa y'}_{mn}|^2+\cev\nu_{z_{mn}^\kappa}|\hat h^{\kappa z'}_{mn}|^2}{|\hat h^{\kappa x'}_{mn}|^2}.
\end{eqnarray}}

{Then, the forward message from $x^t_n$ to $f_{x^t_n}$ can be expressed as
\begin{eqnarray}
m_{x^t_n\to f_{x^t_n}}(x^t_n)&=&\prod_{m,\kappa} m_{f_{h_{mn}^\kappa}\to x^t_n}(x^t_n)\nonumber\\
&=&\CN(x^t_n, \vec x^t_n, \vec \nu_{x^t_n}), \label{eq:msg_u2fu}
\end{eqnarray}
where $\vec\nu_{x^t_n}=1/(\sum_{m,\kappa} 1/{\vec\nu^{\kappa}_{x_{mn}}})$ and $\vec x^t_n=\vec\nu_{x^t_n}\sum_{m,\kappa} \vec x_{mn}^\kappa/{\vec\nu_{x_{mn}}^\kappa}$.
}
So the message $m_{f_{x^t_n}\to x^t_1}(x^t_1)$ can be computed as
\begin{eqnarray}
m_{f_{x^t_n}\to x^t_1}(x^t_1)&=&\int f_{x^t_n} m_{x^t_n\to f_{x^t_n}}(x^t_n) \text{d}{x_n^t}\nonumber\\
&=&\CN\left(x^t_1;\vec x_n, \vec\nu_{x_n}\right), \label{eq:msg_fu2x}
\end{eqnarray}
where $\vec x_n=\vec x^t_n- w_n^x$ and $\vec\nu_{x_n}=\vec\nu_{x_{n}^t}$.  
{Then, the belief of $x^t_1$ can be expressed as
\begin{eqnarray}
b(x^t_1)=\frac{\prod_n m_{f_{x^t_n}\to x^t_1}(x^t_1)}{\int \prod_n m_{f_{x^t_n}(x)\to x^t_1}(x^t_1) \text{d}x^t_1} =\N(x^t_1,\hat x^t_1, \nu_{x^t_1}),\label{eq:blf_x}
\end{eqnarray}
where
$\nu_{x^t_1}={1}/({\sum_n {1}/{\vec\nu_{x_n}}})$ and $\hat x^t_1=\nu_{x^t_1} \sum_n {\vec x_n}/{\vec\nu_{x_n}}$.}

\subsection{{ Backward Message Passing}}
With belief propagation, the backward message $m_{x^t_1\to f_{x_n^t}}(x^t_1)$ can be computed as
\begin{eqnarray}
m_{x^t_1\to f_{x_n^t}}(x^t_1)=\frac{b(x^t_1)}{m_{f_{x_n^t}\to x^t_1}(x^t_1)}=\N(x^t_1;\cev x_n, \cev\nu_{x_n}),\label{eq:msg_x2fx}
\end{eqnarray}
where $\cev\nu_{x_n}=1/(1/\nu_{x^t_1} -1/\vec\nu_{x^t_1})$ and $\cev x_n=\cev\nu_{x_n}(\hat x^t_1/\nu_{x^t_1} -\vec x_n/\vec\nu_{x_n})$.
Then the backward message $m_{f_{x_n^t}\to x_n^t}(x_n^t)$ can be updated by
\begin{eqnarray}
m_{f_{x_n^t}\to x_n^t}(x_n^t)&=&\int f_{x_n^t} m_{x\to f_{u_{x_n}}}(x_n^t)\text{d}x_n^t\nonumber\\
&=&\N(x_n^t; \cev x_n^t, \cev\nu_{x_n^t}),\label{eq:msg_fu2u}
\end{eqnarray}
{where $\cev x_n^t=\cev x_n+w_n^x$ and $\cev\nu_{x_n^t}=\cev\nu_{x_n}$.}
So the belief of $x_n^t$ can be computed as
\begin{align}
b(x_n^t)&=\frac{m_{f_{x_n^t}\to x_n^t}(x_n^t) m_{x_n^t\to f_{x_n^t}}(x_n^t)}{\int m_{f_{x^t_n}\to x^t_n}(x^t_n) m_{x^t_n\to f_{x^t_n}}(x^t_n)\text{d}x^t_n}\nonumber\\
&=\N(x^t_n; \hat x^t_n, \nu_{x^t_n}),\label{eq:blf_u}
\end{align}
with mean $\hat x^t_n$ and variance $\nu_{x^t_n}$ given as
\begin{align}
\nu_{x^t_n}&=1/(1/\cev\nu_{x^t_n}+1/\vec\nu_{x^t_n}),\\
\hat x^t_n&=\nu_{x^t_n}(\cev x^t_n/\cev\nu_{x^t_n} +\vec x^t_n/\vec\nu_{x^t_n} ).
\end{align}

Based on $b(x^t_n)$, we have
\begin{align}
m_{x^t_n\to f_{h_{mn}^\kappa}}(x^t_n)&=\frac{b(x^t_n)}{m_{f_{h_{mn}^\kappa}\to x^t_n}(x^t_n)}\nonumber\\
&=\N(x^t_n; \cev x_{mn}^\kappa, \cev\nu_{x_{mn}}^\kappa),\label{eq:msg_u2fh}
\end{align}
where mean $\cev x_{mn}^\kappa$ and variance $\cev\nu_{x_{mn}}^\kappa$ are given by
\begin{align}
\cev\nu_{x_{mn}}^\kappa&=1/(1/\nu_{x^t_n} -1/\vec\nu_{x_{mn}}^\kappa),\\
\cev x_{mn}^\kappa&=\cev\nu_{x_{mn}}^\kappa(\hat x^t_n/ \nu_{x^t_n} -\vec x_{mn}^\kappa/\vec\nu_{x_{mn}}^\kappa).
\end{align}
Then, we can update the message from $f_{h_{mn}}^\kappa$ to $h_{mn}^\kappa$ by
\begin{align}
&m_{f_{h_{mn}^\kappa}\to h_{mn}^\kappa}(h_{mn}^\kappa) =\int f_{h_{mn}^\kappa} m_{x^t_n\to f_{h_{mn}^\kappa}}(x^t_n)\nonumber\\
&\ \ \  \times  m_{y^t_n\to f_{h_{mn}^\kappa}}(y^t_n) m_{z^t_n\to f_{h_{mn}^\kappa}}(z^t_n)
\text{d}x^t_n\text{d}y^t_n\text{d}z^t_n\nonumber\\
&\ \ \ =\N(h_{mn}; \cev h_{mn}^\kappa, \cev\nu_{h_{mn}}^\kappa), \label{eq:msg_fh2h}
\end{align}
where
\begin{eqnarray}
\cev\nu_{h_{mn}}^\kappa=\xi^\kappa_{mn}+\cev x_{mn}^\kappa \hat h^{\kappa x'}_{mn}+\cev y_{mn}^\kappa \hat h^{\kappa y'}_{mn}+\cev z_{mn}^\kappa\hat h^{\kappa z'}_{mn},\\
\cev h_{mn}^\kappa=\cev \nu_{x_{mn}}^\kappa|\hat h^{\kappa x'}_{mn}|^2+\cev\nu_{y_{mn}}^\kappa|\hat h^{\kappa y'}_{mn}|^2+\cev\nu_{z_{mn}}^\kappa|\hat h^{\kappa z'}_{mn}|^2. \label{eq:prior_h}
\end{eqnarray}

Then we can obtain the belief of channel component $h_{mn}^\kappa$ as
\begin{eqnarray}
b(h_{mn}^\kappa)&=&m_{f_{h_{mn}^\kappa}\to h_{mn}^\kappa}(h_{mn}^\kappa)m_{h^{\kappa}_{mn}\to f_{h^{\kappa}_{mn}}}(h^{\kappa}_{mn})\nonumber\\
&=&\CN(h_{mn}^\kappa;\hat h_{mn}^\kappa,\nu_{h_{mn}^\kappa}) \label{eq:blf_h}
\end{eqnarray}
where 
\begin{eqnarray}
\nu_{h_{mn}^\kappa}&=&1/(1/\cev\nu_{h_{mn}}^\kappa+1/\nu_{q^{\kappa}_{mn}}) \nonumber\\
\hat h_{mn}^\kappa&=&\nu_{h_{mn}^\kappa}(\cev h_{mn}^\kappa/\cev\nu_{h_{mn}}^\kappa+q^{\kappa}_{mn}/\nu_{q^{\kappa}_{mn}}).\label{eq:hatH}
\end{eqnarray}
We stack $\nu_{h_{mn}^\kappa}$ and $\hat h_{mn}^\kappa$, $\forall m,n,\kappa$ as matrices $\bV_H$ and $\hat\bH$. This is the end of backward message passing.

The message passing algorithm is summarized in Algorithm \ref{alg:CE}, {which is called NNHMP (NN-assisted hybrid model based message passing)}. The iteration can be terminated when it reaches the preset maximum number of iterations or the difference between the estimates of two consecutive iterations is less than a threshold.
\begin{algorithm}
\caption{NNHMP for Parametric Channel Estimation}
{\textbf{Initialization}: $\hat\bH=\boldsymbol{0}_{6N\times M}$, $\bV_H=\boldsymbol{1}_{6N\times M}$, $\bS_H=\boldsymbol{0}_{3L\times M}$, $\kappa \in \{xx,xy,xz,yy,yz,zz\}$.\\
\textbf{Repeat}
\begin{algorithmic}[1]
\STATE $\bV_{P}=|\bPhi|^{.2} \bV_H $
\STATE $\bP=\bPhi\hat\bH-\bV_{P}\cdot \bS_H$
\STATE $\bV_{S_H}=1./(\bV_{P}+\gamma^{-1})$
\STATE $\bS_H=\bV_{S_H}\cdot (\bR-\bP)$
\STATE $\bV_{Q_H}=1./(|\bPhi^H|^{.2}\bV_{S_H})$
\STATE $\bQ_H=\hat\bH+\bV_{Q_H}\cdot(\bPhi^H\bS_H)$
\STATE {get $\gamma$ by \eqref{eq:gamma}}  
\STATE $\forall n,m,\kappa$: Compute $\hat h^{\kappa x'}_{mn}, \hat h^{\kappa y'}_{mn}, \hat h^{\kappa z'}_{mn}, \xi^\kappa_{mn}$ with \eqref{eq:nn_approx}
\STATE $\forall n,m,\kappa$: Compute  $m_{f_{h_{mn}}^\kappa\to x_n^t}$ with \eqref{eq:msg_h2u}
\STATE $\forall n$: Compute  $m_{x_n^t\to f_{x_n^t}}$ with \eqref{eq:msg_u2fu}
\STATE $\forall n$: Compute  $m_{f_{x_n^t}\to x_1^t}$ with \eqref{eq:msg_fu2x}
\STATE Compute belief $b(x_1^t)$ with \eqref{eq:blf_x}
\STATE $\forall n$: Compute $m_{f_{x_n^t}\to x_n^t}$ with \eqref{eq:msg_fu2u}
\STATE $\forall n$: Compute $m_{x_1^t\to f_{x_n^t}}$ with \eqref{eq:msg_x2fx}
\STATE $\forall n$: Compute belief $b(x_n^t)$ with \eqref{eq:blf_u}
\STATE $\forall n,m,\kappa$: get $m_{x_n^t\to f_{h_{mn}^\kappa}}$ with \eqref{eq:msg_u2fh}
\STATE Compute $m_{y_n^t\to f_{h_{mn}^\kappa}}$ and $m_{z_n^t\to f_{h_{mn}^\kappa}}$ with the procedure similar to Lines 10-17
\STATE $\forall n,m,\kappa$: Compute $m_{f_{h_{mn}^\kappa}\to h_{mn}^\kappa}$ with \eqref{eq:msg_fh2h}
\STATE $\forall n,m,\kappa$: get $b(h_{mn}^\kappa)=\CN(h_{mn}^\kappa;\hat h_{mn}^\kappa,\nu_{h_{mn}^\kappa})$ by \eqref{eq:blf_h}
\STATE Stack $\hat h_{mn}^\kappa, \nu_{h_{mn}^\kappa}, \forall n,m,\kappa$ into $\hat\bH$ and $\bV_H$
\end{algorithmic}
\textbf{Until terminated}
\label{alg:CE}}
\end{algorithm}
{We can see that the NNHMP algorithm includes three parts: UAMP part (Lines 1-7), the part related to the NN-assisted hybrid local function node (Line 9) and the part producing location estimation (Lines 9-19). The UAMP part is dominated by matrix multiplication with a complexity of $\mathcal{O}(MNL)$. 
In the part related to the NN node, in total $MN$ channel elements are involved, and a multiplication of two matrices with dimensions $3\times N_h$ and $N_h\times 12$ is performed, so the complexity is $\mathcal{O}(MNN_h^2)$, where $N_h$ denotes the number of hidden nodes. In the remaining part, the highest complexity lies in the computation of message $m_{f_{h_{nm}^\kappa}\to x_n^t}(x_n^t), \forall n,\kappa$ in \eqref{eq:msg_fu2x}, whose complexity is $\mathcal{O}(MN)$. 

\begin{figure}[!t]
\centering
\includegraphics[width=0.4\textwidth]{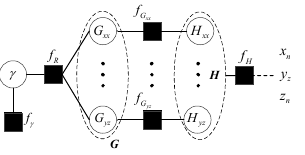}
\centering
\caption{Factor graph representation of \eqref{eq:factorizationRF}.}
\label{fig:FactorGraph2}
\end{figure}

\section{Channel Estimation with Hybrid Receiver} \label{RFChains}
In the previous section, we assume a full digital receiver, where each antenna patch is connected to a RF chain \cite{Demir2022WCL,Luca2023arXiv}. In this section, we consider a more practical hybrid structure \cite{yu2023bayes,zhang2019hybrid}, where the holographic surface is connected to $P$ ( $P < M$) RF chains. This leads to a $3P \times 3M$ matrix $\tilde\bF$ in the signal model, i.e., the received signal can be expressed as 
\begin{eqnarray}
\tilde\bY=\tilde\bF\tilde\bH\tilde\bS+\tilde\bW, 	\label{RFmodel}
\end{eqnarray}
where $\tilde\bY \triangleq [\tilde \bY_x\tra,\tilde \bY_y\tra,\tilde \bY_z\tra]\tra \in \mathbb{C}^{3P\times L}$ with $\tilde \bY_x, \tilde \bY_y$ and $\tilde \bY_z\in \mathbb{C}^{P\times L}$, and $\tilde\bS$, $\tilde\bH$ and $\tilde\bW$ are the same as those in \eqref{eq:Recv0}.
To facilitate channel estimation, the model is rewritten as 
\begin{eqnarray}
\bY&=&
\bS
\begin{bmatrix}
(\bF \tilde \bH_{xx})\tra \\
\vdots \\
(\bF \tilde \bH_{yz})\tra 
\end{bmatrix}+\bW =
\bS
\begin{bmatrix}
\bG_{xx} \\
\vdots \\
\bG_{yz} 
\end{bmatrix}+\bW \nonumber\\
&\triangleq& \bS\bG+\bW, \label{eq:Recv4}
\end{eqnarray}
where $\bY \triangleq [\tilde{\boldsymbol{Y}}_x,\tilde{\boldsymbol{Y}}_x,\tilde{\boldsymbol{Y}}_x]\tra \in \mathbb{C}^{3L\times P}$.}
{Define $\bG \triangleq [\bG_{xx}\tra,\cdots,\bG_{yz}\tra]\tra\in \mathbb{C}^{6N\times P}$ and 
\begin{eqnarray}
\bG_{\kappa}\tra=\bF \tilde \bH_{\kappa}=\bF\bH_\kappa\tra\in \mathbb{C}^{P\times N}, 
\end{eqnarray}
where $\bH_\kappa$ has the same definition in \eqref{eq:Recv2}. Comparing it with model \eqref{eq:Recv3}, we can see that the difference lies in the likelihood function $p(\bR|\bH,\lambda)$, which can be expressed as 
\begin{eqnarray}\label{eq:factorizationRF}
&p(\bR|\bG, \bH,\lambda)=p(\bR|\bG,\gamma) \prod_{\kappa}
p(\bG_\kappa|\bH_\kappa) p(\gamma)\nonumber\\
&\ \ \ \  =f_R(\bR,\bG,\gamma)\prod_{\kappa}f_{G_\kappa}(\bG_\kappa,\bH_\kappa)p(\gamma)
\end{eqnarray}
where
$p(\bR|\bG,\gamma)=\CN(\bR;\boldsymbol{\Phi}\bG,\gamma^{-1}\bI)$ and $p(\bG_{\kappa}|\bH_{\kappa})=\delta(\bG_{\kappa}\tra-\bF\bH_{\kappa}\tra)$ with $\bPhi$ and $\bR$ defined in \eqref{eq:factorization}.

Part of the graph representation in this case is shown in Fig.\ref{fig:FactorGraph2}, and the remaining part of the graph representation is the same as that in Fig.\ref{fig:FactorGraph}. Hence, we focus on message passing in Fig.\ref{fig:FactorGraph2}  in this section.  
It can be seen from the factorization \eqref{eq:factorizationRF} that there are two successive linear mixing operations involved in the signal model, which are $f_{\bR}(\bR,\bG,\gamma)$ and $f_{\bG_{\kappa}}(\bG_{\kappa},\bH_{\kappa})$. To solve the problem, we propose using the cascade of two UAMP algorithms.

\subsection{UAMP for $f_{\bR}(\bR,\bG,\gamma)$}
We can see from \eqref{eq:factorizationRF} that $p(\bR|\bG,\gamma)$ has essentially the same expression as $p(\bR|\bH,\gamma)$ in \eqref{eq:factorization}. Similar to \eqref{eq:PH}-\eqref{eq:VQH}, we have the following steps according to UAMP. 

We first compute matrices $\bV_{P_G}$ and $\bP_G$ by
\begin{eqnarray}
\bV_{\bP_G}&=&|\bPhi|^{.2} \bV_G, \nonumber\\
\bP_G&=&\bPhi \hat\bG-\bV_{P_G} \cdot \bS_G, \nonumber
\end{eqnarray}	
where ${\bPhi=\bLambda_G\bV_G}$ and $[\bU_G,\Lambda_G,\bV_G]=\text{SVD}(\bS)$.
The estimate of the noise precision is updated as 
\begin{eqnarray}
{\hat\gamma}=\frac{3PL}{||\bR-\bZ_G||^2+\bV_{Z_G}}, \label{eq:gamma2}
\end{eqnarray}
where 
\begin{eqnarray}
\bV_{Z_G}&=&1./(\hat\gamma+1./\bV_{P_G}), \nonumber\\
\bZ_G&=&\bV_{Z_G}\cdot(\hat\gamma \bR+\bP_G./\bV_{P_G}). \nonumber
\end{eqnarray}
Then the  two intermediate matrices can be computed by 
\begin{eqnarray}
\bV_{S_G}&=&1./(\bV_{P_G}+\hat\gamma^{-1}),\nonumber\\
\bS_G&=&\bV_{S_G}\cdot (\bR-\hat\bP_G). \nonumber
\end{eqnarray}
After that we can get $\bV_{Q_G}$ and $\bQ_G$ as
\begin{align}
\bV_{Q_G}&=1./(|\bPhi\herm|^{.2}\bV_{S_G}),\label{eq:VQE}\\
\bQ_G&=\hat\bG+\bV_{Q_G}\cdot(\bPhi\herm\bS_G), \label{eq:QE}
\end{align}
They can be represented using block matrices as
\begin{eqnarray}
\bV_{Q_G}=
\begin{bmatrix}
\bV_{Q_{G}^{xx}}\\
\vdots \\
\bV_{Q_{G}^{yz}}
\end{bmatrix},
\ \ \ \bQ_G=
\begin{bmatrix}
\bQ_{G_{xx}}\\
\vdots \\
\bQ_{G_{yz}}
\end{bmatrix}.
\end{eqnarray} 
where $\bV_{Q_{G}^{\kappa}}\in\mathbb{R}^{N\times P}$ and $\bQ_{G_{\kappa}}\in\mathbb{C}^{N\times P}$ with their $(n,m)$-th element being $\vec g_{mn}^\kappa$ and $\vec\nu_{g_{mn}^\kappa}$. They provide the mean and variance of message $m_{g_{mn}^\kappa\to f_{g_{mn}^\kappa}}(g_{mn}^\kappa)$, i.e., $m_{g_{mn}^\kappa\to f_{g_{mn}^\kappa}}(g_{mn}^\kappa)=\CN(g_{mn}^\kappa; \vec g_{mn}^\kappa, \vec\nu_{g_{mn}^\kappa})$.
Then the belief of $g_{mn}^\kappa$ is $b(g_{mn}^\kappa)=\CN(g_{mn}^\kappa;\hat g_{mn}^\kappa,\nu_{g_{mn}^\kappa})$, where
\begin{eqnarray}
\nu_{g_{mn}^\kappa}&=&1/(1/\vec\nu_{g_{mn}^\kappa}+1/\nu_{p_{h_{mn}^\kappa}})\nonumber\\
\hat g_{mn}^\kappa&=&\nu_{g_{mn}^\kappa}(\vec g_{mn}^\kappa/\vec\nu_{g_{mn}^\kappa}+p_{h_{mn}^\kappa}/\nu_{p_{h_{mn}^\kappa}})\label{eq:blf_g}
\end{eqnarray}
with $p_{h_{mn}^\kappa}$ and $\nu_{p_{h_{mn}^\kappa}}$ being the $(n,m)$-th member of $\bP_{H_{\kappa}}$ and $\bV_{P_{H_{\kappa}}}$, and they are updated in \eqref{eq:VPH} and \eqref{eq:PH}. Next, we stack $\hat g_{mn}^\kappa$ and $\nu_{g_{mn}^\kappa}, \forall m,n,\kappa$ into matrices $\hat\bG$ and $\bV_G$. Collectively, we have matrices 
\begin{eqnarray}
\bV_{G_\kappa}&=&1./(1./\bV_{Q_{G}^{\kappa}}+1./\bV_{P_{H_{\kappa}}})\nonumber\\	\hat\bG&=&\bV_{G_\kappa}\cdot(\bQ_{G_{\kappa}}./\bV_{Q_{G}^{\kappa}}+\bP_{H_{\kappa}}./\bV_{P_{H_{\kappa}}}).\nonumber
\end{eqnarray}

\subsection{UAMP for $f_{\bG_{\kappa}}(\bG_{\kappa},\bH_{\kappa})$}
As $\bG_{\kappa}\tra=\bF\bH_{\kappa}\tra$, we can construct a pseudo-observation model of $\bH_{\kappa}\tra$, i.e., 
\begin{eqnarray}
\bQ_{G_\kappa}\tra=\bF \bH_{\kappa}\tra+ \bW_{G_\kappa}, \label{eq:pseu_G}
\end{eqnarray}	
where $\bW_{G_\kappa}\in\mathbb{C}^{N\times P}$ denotes a white Gaussian noise matrix, and the variances of each element is given by the elements in $\bV_{G_\kappa}\tra$.} To use the UAMP algorithm, we transform \eqref{eq:pseu_G} into
\begin{eqnarray}
\bR_{G_\kappa}\tra=\bU_F\herm\bQ_{G_\kappa}\tra=\bPhi_H \bH_{\kappa}\tra+ \bW_{G_\kappa}, \nonumber
\end{eqnarray}	
where $[\bU_F,\Lambda_F,\bV_F]=\text{SVD}(\bF)$, $\bR_{G_\kappa}\tra=\bU_F\herm\bQ_{G_\kappa}\tra\in \mathbb{C}^{P\times N}$ and $\bPhi_H=\Lambda_H\bV_H\in \mathbb{C}^{P\times M}$.
Then, according to UAMP, two auxiliary matrices $\bV_{P_{H_{\kappa}}}$ and $\bP_{H_{\kappa}}$ can be computed by
\begin{align}
&\bV_{P_{H_{\kappa}}}={|\bPhi_H|}^{2}\bV\tra_{\bH_{\kappa}}, \label{eq:PH}\\
&\bP_{H_{\kappa}}=\bPhi_H \hat\bH\tra_{\kappa}-\bV_{P_{H_{\kappa}}}\cdot  \bS_{H_{\kappa}},\label{eq:VPH}
\end{align}
with which intermediate matrices $\bV_{S_{H_{\kappa}}}$ and $ \bS_{H_{\kappa}}$ are updated as
\begin{eqnarray}
&\bV_{S_{H_{\kappa}}}=1./(\bV_{G_{\kappa}}\tra+\bV_{P_{H_\kappa}}),\nonumber\\
& \bS_{H_{\kappa}}=\bV_{S_{H_{\kappa}}}\cdot(\bR_{G_\kappa}\tra-\bP_{H_\kappa}). \label{eq:SH}
\end{eqnarray}	
Then, we can compute matrices $\bQ_{H_{\kappa}}$ and $\bV_{\bQ_{H_{\kappa}}}$ with
\begin{align}
&\bV_{\bQ_{H_{\kappa}}}\tra=1./({|\bPhi_{\bH}^{\text{H}}|}^{2}\bV_{S_{H_{\kappa}}}), \label{eq:QH}\\
&\bQ_{H_{\kappa}}\tra= \hat \bH_{\kappa}\tra+ \bV_{\bQ_{H_{\kappa}}} \cdot (\bPhi_{\bH}^{\text{H}} \bS_{H_{\kappa}} ).\label{eq:VQH}
\end{align}
The $(n,m)$-th elements of $\bV_{\bQ_{H_{\kappa}}}$ and $\bQ_{H_{\kappa}}$ 
denoted as $q_{mn}^{\kappa}$ and $\nu_{q_{mn}^{\kappa}}$ represent the mean and variance of extrinsic message $m_{h^{\kappa}_{mn}\to f_{h^{\kappa}_{mn}}}(h^{\kappa}_{mn})$. Here message $m_{h^{\kappa}_{mn}\to f_{h^{\kappa}_{mn}}}(h^{\kappa}_{mn})$ corresponded to the message \eqref{eq:msg_q} in the full digital case. So we can use \eqref{eq:nn_approx}-\eqref{eq:blf_h} to get the belief $b(h_{mn}^\kappa)=\CN(h_{mn}^\kappa;\hat h_{mn}^\kappa,\nu_{h_{mn}^\kappa})$. 
Stacking $\hat h_{mn}^\kappa$ and $\nu_{h_{mn}^\kappa}, \forall m,n,\kappa$ into matrices $\hat\bH, \bV_{H}$ completes the message passing.
The message passing procedure is summarized in Algorithm \ref{alg:CE2}.

\begin{algorithm}
\caption{NNHMP Algorithm for Hybrid Receiver}
\textbf{Initialization}: $\hat{\bH}_{\kappa}=\boldsymbol{0}_{N\times M} $,  $\bV_{\bH_{\kappa}}=\boldsymbol{1}_{N \times M}$, $ \bS_{H_{\kappa}}=\boldsymbol{0}_{P\times N}$, $\bV_{G}=\boldsymbol{1}_{6N\times P}$,  $\hat\bG=\boldsymbol{0}_{6N\times P}$, $ \hat\bS_G=\boldsymbol{0}_{3L\times P}$, $\hat\gamma=1$, $\kappa \in \{xx,xy,xz,yy,yz,zz\}$. \\
\textbf{Repeat}
\begin{algorithmic}[1]
\STATE  $\bV_{P_G}=|\bPhi|^{.2} \bV_G$
\STATE  $\bP_G=\bPhi \hat\bG-\bV_{P_G}\cdot\hat\bS_G$
\STATE	$\bV_{Z_G}=1./(\hat\gamma+1./\bV_{P_G})$
\STATE  $\bZ_G=\bV_{Z_G}\cdot(\hat\gamma \bR+\bP_G./\bV_{P_G})$
\STATE Compute noise precision $\hat\gamma$ by \eqref{eq:gamma2}
\STATE $\bV_{S_G}=1./(\bV_{P_G}+\hat\gamma^{-1})$
\STATE  $\bS_G=\bV_{S_G}\cdot (\bR-\hat\bP_G)$
\STATE $\bV_{Q_G}=1./(|\bPhi\herm|^{.2}\bV_{S_G})$
\STATE $\bQ_G=\hat\bG+\bV_{Q_G}\cdot(\bPhi\herm\bS_G)$
\STATE Obtain $\hat g_{mn}^\kappa$, and $\nu_{g_{mn}^\kappa}, \forall m,n,\kappa$ by \eqref{eq:blf_g}, stack them into $\hat\bG$ and $\bV_G$
\STATE $\forall \kappa, \bR_{G_\kappa}\tra=\bU_F\herm\bQ_{G_\kappa}\tra$
\STATE $\forall \kappa, \bV_{P_{H_{\kappa}}}={|\bPhi_H|}^{2}\bV\tra_{\bH_{\kappa}}$
\STATE $\forall \kappa, \bP_{H_{\kappa}}=\bPhi_H \hat\bH\tra_{\kappa}-\bV_{P_{H_{\kappa}}}\cdot  \bS_{H_{\kappa}}$
\STATE $\forall \kappa, \bV_{S_{H_{\kappa}}}=1./(\bV_{G_{\kappa}}\tra+\bV_{P_{H_\kappa}})$
\STATE $\forall \kappa, \bS_{H_{\kappa}}=\bV_{S_{H_{\kappa}}}\cdot(\bR_{G_\kappa}\tra-\bP_{H_\kappa})$
\STATE $\forall \kappa, \bV_{\bQ_{H_{\kappa}}}\tra=1./({|\bPhi_{\bH}^{\text{H}}|}^{2}\bV_{S_{H_{\kappa}}})$
\STATE $\forall \kappa, \bQ_{H_{\kappa}}\tra= \hat \bH_{\kappa}\tra+ \bV_{\bQ_{H_{\kappa}}} \cdot (\bPhi_{\bH}^{\text{H}} \bS_{H_{\kappa}} )$
\STATE Obtain $\hat\bH_{\kappa}, \bV_{\bH_{\kappa}}, \forall \kappa$ by lines $8-20$ of algorithm \ref{alg:CE} 
\end{algorithmic}
\textbf{Until terminated}
\label{alg:CE2}
\end{algorithm}

\section{Simulation Results}
In this section, we provide extensive numerical results to demonstrate the performance of the proposed method. The system settings are as follows. 
The carrier frequency is set to $f=3$GHz. At the base station, we assume a surface with $10\times 10$ antenna patches, i.e., $M=100$. 
At the user side, the surface consists of $5\times 5$ patches, i.e., $N=25$. The patch sizes of the base station and the user are set to  $\Delta_x^r=\Delta_y^r=0.05$m and $\Delta_x^t=\Delta_y^t= 0.01$m respectively. 
The modulation scheme used is QPSK, and the pilot signals are QPSK symbols, which are randomly generated. {We vary the number of received signal vectors $L$ from 100 to 500.} 
As mentioned before, the number of hidden nodes in the NN hidden layer $N_h=50$. 
The coordinate of the first patch of the base station is  $(x_1^r,y_1^r,z_1^r)=(0,0,0)$, and the coordinate of the first patch of the user is randomly generated with $x_1^t \sim \text{U}[-1,1]\text{m}, y_1^t \sim \text{U}[-1,1]\text{m}, z_1^t \sim \text{U}[20,40]\text{m}$. 
We evaluate the performance of estimators in terms of the normalized mean squared error of the channel and the location of the user (if an estimator provides the estimate of the user location), i.e., $\text{NMSE}_H$ and $\text{NMSE}_p$, 
which are defined as
\begin{eqnarray}
&&\text{NMSE}_H=E \frac{\|\bH-\hat\bH\|^2}{\|\bH\|^2}\nonumber\\
&&\text{NMSE}_p=E \frac{\|[x_1^t,y_1^t,z_1^t]^T-[\hat x_1^t,\hat y_1^t,\hat z_1^t]^T\|^2}{\|[x_1^t,y_1^t,z_1^t]^T\|^2}\nonumber
\end{eqnarray}
where $(x_1^t,y_1^t,z_1^t)$ and $\bH$ represent the user coordinate and the channel matrix, 
and $(\hat x_1^t,\hat y_1^t,\hat z_1^t)$ and $\hat\bH$ denote their estimates.

To the best of our knowledge, there are no existing works on parametric HMIMO channel estimation in the literature. For comparison, we include the performance of the LS channel estimation based on \eqref{eq:Recv2}, which is given as
\begin{eqnarray}
\hat\bH_{LS}=(\bS^H\bS)^{-1}\bS^H\bY, \nonumber
\end{eqnarray} 
where the channel matrix is directly estimated, and the estimate of the user location cannot be provided. 
In addition, it is noted that the approximate channel model \eqref{Appmod} is still complex. We can also develop an NN-assisted hybrid model to replace it, so that a message passing algorithm can be also developed to achieve parametric channel estimation. The related simulation results are indicated by "AppMod". We will show that the approximate model can lead to considerable performance loss due to its significant mismatch with the actual channel model. In addition, we also include two performance bounds. One bound is obtained by assume the coordinate of the user, i.e., $(x_1^t,y_1^t,z_1^t)$ is known, so that the channel matrix can be constructed using \eqref{eq:Hmn_NN}. Moreover, the CRLB of user location estimation is also included, and the derivation is shown in Appendix.

\begin{figure}[!t]
\centering
\includegraphics[width=0.45\textwidth]{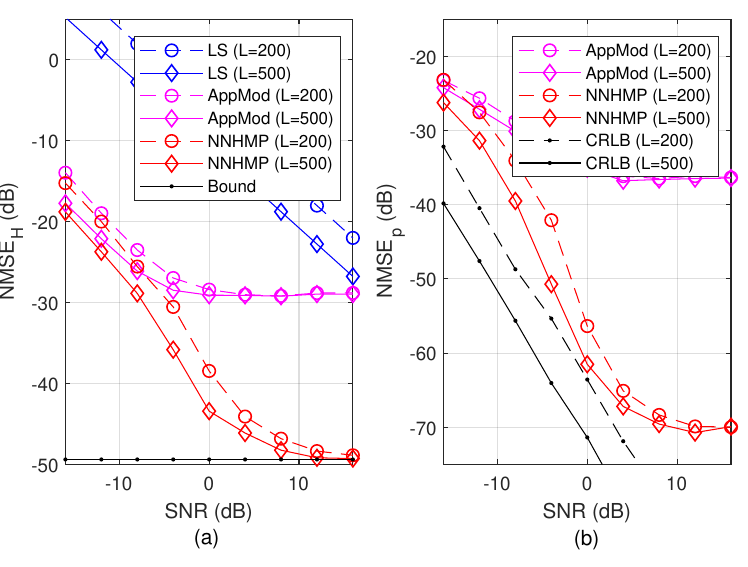}
\centering
\caption{NMSE performance of the estimators versus SNR.}
\label{fig:MSEvsSNR5001}
\end{figure}

The NMSE performance of the estimators versus SNR with $L=200$ and $500$ are shown in Fig. \ref{fig:MSEvsSNR5001}, where (a) and (b) show NMSE$_H$ and NMSE$_p$, respectively. It can be seen from Fig. \ref{fig:MSEvsSNR5001} (a) that, with the increase of the SNR, the performance of the proposed method gradually approaches the bound with perfect user location. This is because the estimation accuracy of the user location is improved with the SNR as shown in Fig. \ref{fig:MSEvsSNR5001} (b). As expected, the proposed method significantly outperforms the LS one due to the strategy of parametric estimation. We can see from Fig. \ref{fig:MSEvsSNR5001} (b) that the performance of proposed method delivers performance close to the CRLB at low SNRs, while deviates from the CRLB at high SNRs due to the small model mismatch.  In both Figs. \ref{fig:MSEvsSNR5001} (a) and (b), we can also see that the proposed method delivers considerably better performance than AppMod, as the true channels can be well characterized by the proposed hybrid channel model. Due to the considerable model mismatch, AppMod exhibits a high error floor. For the proposed method, we observe a much lower error floor. This is because the NN with limited number of hidden nodes cannot perfectly model the channel.



\begin{figure}[!t]
\centering
\includegraphics[width=0.45\textwidth]{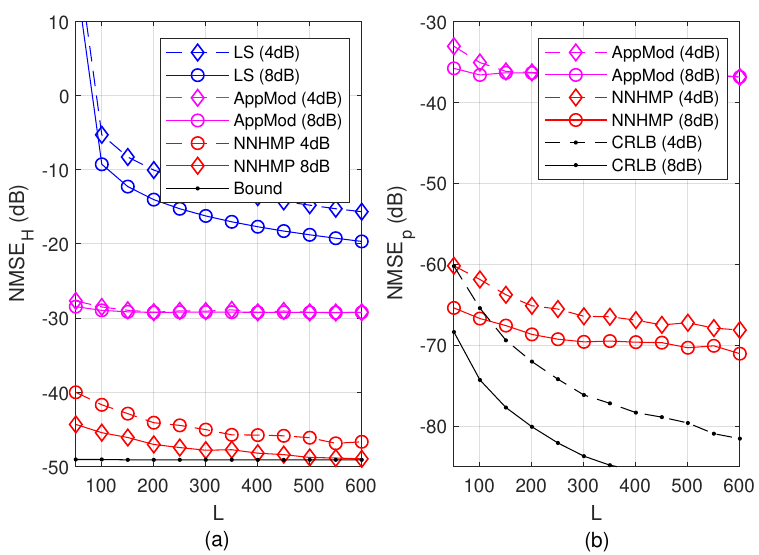}
\centering
\caption{NMSE performance of the estimators versus pilot length $L$ .}
\label{fig:MSEvsL}
\end{figure}

\begin{figure}[!t]
\centering
\includegraphics[width=0.45\textwidth]{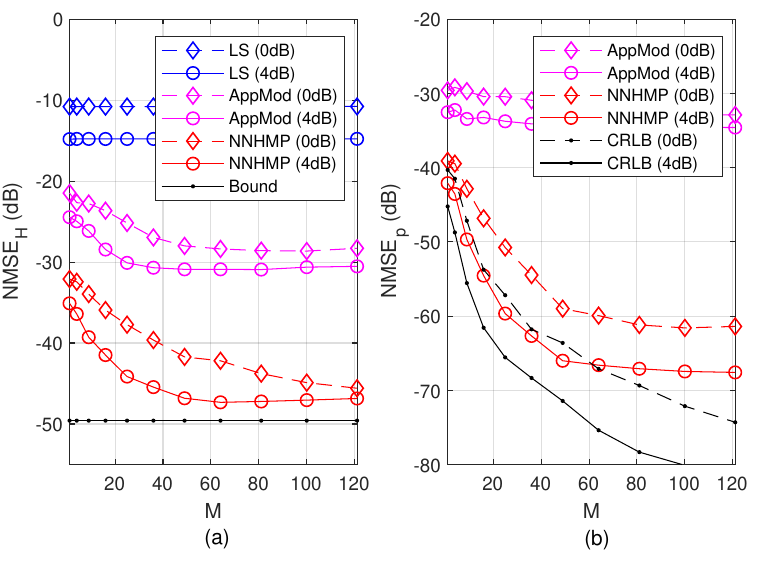}
\centering
\caption{NMSE performance of the estimators versus $M$.}
\label{fig:MSEvsM}
\end{figure}
Then we examine the performance of the methods versus the number of received signal vectors (i.e., the length of pilot signal) $L$ in Fig. \ref{fig:MSEvsL}, where the SNR is set to 4dB and 8dB. As expected, with the increase of $L$, the performance of the LS and the proposed estimator improve. However, AppMod has a high error floor due to the significant model mismatch. We can see that the proposed estimator delivers significantly better performance and it approaches the performance bound when  $L$ is relatively large in  Fig. \ref{fig:MSEvsL}(a). In Fig. \ref{fig:MSEvsL}(b), we can observe significant gaps between the performance of the proposed method and the CRLB. This is again because the small model mismatch of the proposed method dominates the error performance when the NMSE is very small (e.g., less than -70dB), resulting in an error floor.  

In Fig. \ref{fig:MSEvsM}, we examine the estimation performance $\text{NMSE}_H$ and $\text{NMSE}_p$ versus the number of BS antenna patches $M$ at various SNRs. With the increase of $M$, the NMSE performance of the parametric estimators improves until performance floors appear. Compared to AppMod, the proposed estimator has a much lower floor thanks to the high modeling capability of the hybrid channel model. We can also see that the performance of LS channel estimator does not improve. This is because the LS estimator is not a parametric one, and the number of channel coefficients to be estimated also increases with $M$.  Again the gaps between the proposed method and CRLB is due to the small modelling error, which dominates the NMSE performance.   




\begin{figure}[!t]
\centering
\includegraphics[width=0.45\textwidth]{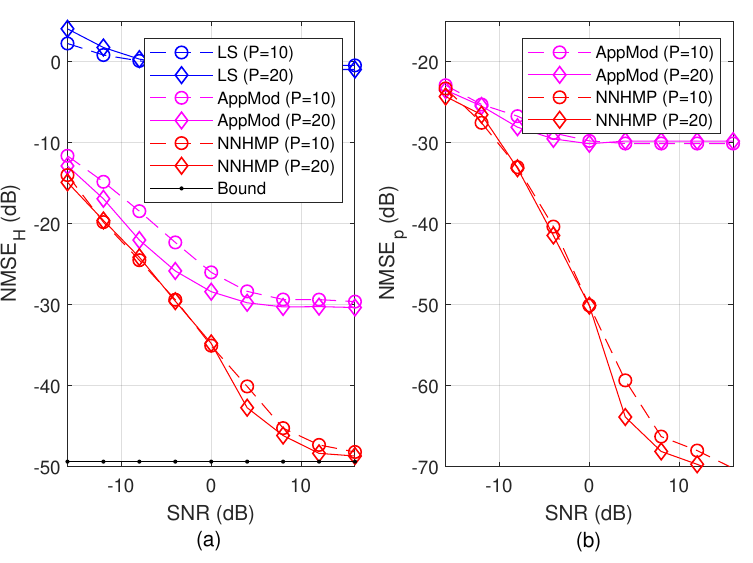}
\centering
\caption{NMSE of the estimators with hybrid receiver versus SNR.} 
\label{fig:MSEvsSNR_RF}
\end{figure}

\begin{figure}[!t]
\centering
\includegraphics[width=0.45\textwidth]{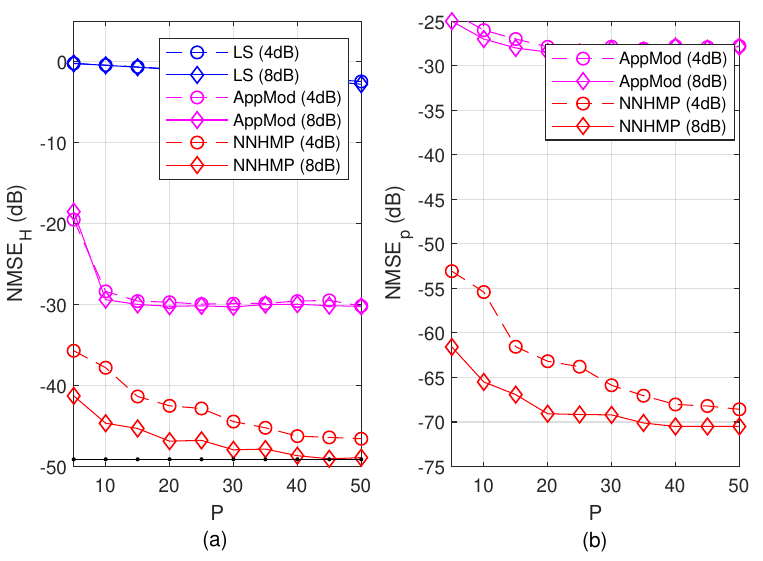}
\centering
\caption{NMSE of the estimators with hybrid receiver versus $P$.} 
\label{fig:MSEvsP_RF}
\end{figure}

Next, we examine the performance of various estimators in the case of a hybrid receiver. The NMSE performance of the estimators versus SNR and the number of RF chains $P$ is shown in Fig.\ref{fig:MSEvsSNR_RF} and \ref{fig:MSEvsP_RF}. From the results, it can be seen that with the increase of the number of RF chains and SNR, the performance of the estimators improves as expected. We again observe that AppMod has a high performance floor in all the cases due to the significant model mismatch and the LS estimator does not perform well as the dimension of the received signal vectors is significantly reduced. The proposed estimator achieves the best performance and performs significantly better than other estimators.  



\section{Conclusions}
In this paper, we have investigated the issue of channel estimation for HMIMO, where the channel is characterized by the dyadic Green's function. Considering that the channel matrix is parameterized with a few parameters, we propose a parametric channel estimation method to achieve superior estimation performance. To tackle the challenging complex nonlinear relationship between the parameters and channel coefficients, we develop a hybrid channel model with the aid of an NN. 
With the hybrid channel model, the estimation problem is formulated in a probabilistic form. Leveraging a factor graph representation and UAMP, an efficient message passing algorithm is developed. Extensive simulation results show the superior performance of the proposed method. 

\appendices
\section{Cramér-Rao Lower Bound Using the Hybrid Channel Model}
It is worth mentioning that, regarding the user position estimation, the hybrid channel model  provides a convenient way to obtain the CRLB thanks to its simpler expression. 

According to \eqref{eq:Recv2}, 
the received signal $\bY=(\by_m,...,\by_M)\in \mathbb{C}^{3L \times M}$ is given as
\begin{eqnarray}
\boldsymbol{Y}=\boldsymbol{S}\boldsymbol{H}+\boldsymbol{W},\nonumber
\end{eqnarray}
where $\bW\in \mathbb{C}^{3L \times M} $ is the AWGN with precision $\gamma$. We assume that the precision is known in the derivation of the CRLB, while it is estimated in the proposed method. Define vector $\bp=[x_1^t,y_1^t,z_1^t]^T \in \mathbb{R}^{3\times 1}$, which includes the unknown coordinate. The logarithm of the likelihood function can be expressed as
\begin{eqnarray}\nonumber
\ln p(\bY|x_1^t,y_1^t,z_1^t)=- \left\| \bY-\bS\bH \right\|^{2} {\gamma}+C \nonumber\\
=-\sum_{m=1}^{M}   \left\| \by_m-\bS\bh_m \right\|^{2} {\gamma}+C \nonumber
\end{eqnarray}
where $C$ is a constant, vectors $\bh_m=[\bh_{mn}\tra,...,\bh_{mN}\tra]\tra$, $\boldsymbol{h}_{mn}=[h^{xx}_{mn}, h^{yy}_{mn}, h^{zz}_{mn}, h^{xy}_{mn}, h^{xz}_{mn}, h^{yz}_{mn}]\tra\in \mathbb{C}^{6 \times 1}$.
Then, we define $f({\bp})=\sum_{m=1}^{M}f_m({\bp})$, where $f_m({\bp})=- \left\| \by_m-\bS\bh_m \right\|^{2}\gamma$. 

The Fisher information matrix (FIM) $\boldsymbol{\mathcal{F}}({\bp})\in \mathbb{R}^{3\times 3}$ can be obtained as
\begin{eqnarray}\nonumber
\boldsymbol{\mathcal{F}}({\bp})
=-\mathbb{E} \left[
\begin{matrix}  
\frac{\partial^2 f(\bp)}{\partial x_1^t\partial x_1^t}  & \frac{\partial^2 f(\bp)}{\partial x_1^t \partial y}  & \frac{\partial^2 f(\bp)}{\partial x_1^t \partial z_1^t}  \\  
\frac{\partial^2 f(\bp)}{\partial y \partial x_1^t}  & \frac{\partial^2 f(\bp)}{\partial y \partial y}  & \frac{\partial^2 f(\bp)}{\partial y \partial z_1^t}  \\  
\frac{\partial^2 f(\bp)}{\partial z_1^t \partial x_1^t}  & \frac{\partial^2 f(\bp)}{\partial z_1^t \partial y}  & \frac{\partial^2 f(\bp)}{\partial z_1^t \partial z_1^t}  
\end{matrix}\right].
\end{eqnarray}
In the following, we only take element $\frac{\partial^2 f(\bp) }{\partial x_1^t \partial y_1^t} $ as example, and derivations for other elements will be the same. We have
\begin{eqnarray}
\frac{\partial^2 f(\bp) }{\partial x_1^t \partial y_1^t} =\sum_m \frac{\partial^2 f_m(\bp) }{\partial x_1^t \partial y_1^t}\nonumber
\end{eqnarray}
where
\begin{align}
&\frac{\partial^2 f_m(\bp) }{\partial x_1^t \partial y_1^t} 
= {\gamma}  \Big( \by_m^H \bS \frac{\partial^2 \bh_m}{\partial x_1^t \partial y_1^t}+\frac{\partial^2 \bh_m^H}{\partial x_1^t \partial y_1^t} \bS^H \by- \frac{\partial^2 \bh_m^H}{\partial x_1^t \partial y_1^t} \bS^H \bS \bh_m \nonumber \\
& -\frac{\partial \bh_m^H}{\partial x_1^t} \bS^H \bS \frac{\partial \bh_m}{\partial y_1^t}-\frac{\partial \bh_m^H}{\partial y_1^t} \bS^H \bS \frac{\partial \bh_m}{\partial x_1^t}-\bh_m^H \bS^H \bS \frac{\partial^2 \bh_m}{\partial x_1^t \partial y_1^t}  \Big),\nonumber
\end{align}
The partial derivation $\frac{\partial \bh_{mn}}{\partial x_1^t}=\left(\frac{\partial h^{xx}_{mn}}{\partial x_1^t},...,\frac{\partial h^{yz}_{mn}}{\partial x_1^t }\right)$ and $\frac{\partial^2 \bh_m}{\partial x_1^t \partial y_1^t}=\left(\frac{\partial^2 h^{xx}_m}{\partial x_1^t \partial y_1^t},...,\frac{\partial^2 h^{yz}_{mn}}{\partial x_1^t \partial y_1^t}\right)$. From \eqref{eq:Hmn_NN} we have $h^{\kappa}_{mn}=\varphi^{\kappa}(x_{mn},y_{mn},z_{mn})\exp(ik_0 r_{mn}), \forall\kappa\in\{xx,...,yz\}$, with 
\begin{align}
&\phi_{mn}^{\kappa}\triangleq\varphi^{\kappa}(x_{mn},y_{mn},z_{mn})\nonumber\\
&\ \ \ \ \ =\varphi^{\kappa}(x_1^t+\Delta_{mn}^x,y_1^t+\Delta_{mn}^y,z_1^t)\nonumber
\end{align}
where $\Delta_{mn}^x$ and  $\Delta_{mn}^y$ are defined in \eqref{eq:coordinate}, representing the offset of $(m,n)$-th patch pair in the $x,y$ directions, respectively. 
So the first and second partial derivatives are obtained as
\begin{align}
&\frac{\partial h^{\kappa}_{mn}}{\partial x_1^t}=\left( \frac{\partial\phi^{\kappa}_{mn}}{\partial x_1^t} +ik_0 \phi^{\kappa}_{mn}\frac{\partial r_{mn}}{\partial x_1^t}\right){\exp(ik_0 r_{mn})}. \nonumber\\
&\frac{\partial h^{\kappa}_{mn}}{\partial x_1^t \partial y_1^t}=  \nonumber\\
&\ \ \ \frac{\partial}{\partial y_1^t} \left( \frac{\partial\phi^{\kappa}_{mn}}{\partial x_1^t}\exp(ik_0 r_{mn}) +ik_0\phi^{\kappa}_{mn} \frac{\partial r_{mn}}{\partial x_1^t}\exp(ik_0 r_{mn})\right) \nonumber \\
&\ \ \ = \left( \frac{\partial^2\phi^{\kappa}_{mn}}{\partial x_1^t \partial y_1^t} +ik_0 \left( \frac{\partial\phi^{\kappa}_{mn}}{\partial x_1^t} \frac{\partial r_{mn}}{\partial y_1^t} + \frac{\partial\phi^{\kappa}_{mn}}{\partial y_1^t} \frac{\partial r_{mn}}{\partial x_1^t}  \right. \right. \nonumber \\
& \left. \left. \ \ \ \ \ \ \ +\phi^{\kappa}_{mn} \frac{\partial^2 r_{mn}}{\partial x_1^t \partial y_1^t}+   ik_0\phi^{\kappa}_{mn}\frac{\partial r_{mn}}{\partial x_1^t}\frac{\partial r_{mn}}{\partial y_1^t}\right) \right)  \exp(ik_0 r_{mn}).\nonumber
\end{align}
As $\phi^{\kappa}_{mn}=(\bw_{2,\kappa_1}+\bw_{2,\kappa_1})\tra g_a(x_{mn}\bw_1^x+y_{mn}\bw_1^y+z_{mn}\bw_1^z+\bb_1)$, the first and second order partial derivatives of $\phi^{\kappa}_{mn}$ are given as 
\begin{eqnarray}
\frac{\partial \phi^{\kappa}_{mn}}{\partial x_1^t}
&=&\left((\bw_{2,\kappa_1}+\bw_{2,\kappa_1})\cdot \bw_1^x\right)\tra g'_a(\bc_{mn}),\nonumber\\
\frac{\partial^2 \phi^{\kappa}_{mn}}{\partial x_1^t\partial y_1^t }
&=&\left((\bw_{2,1}+\bw_{2,7})\cdot (\bw_1^x \cdot \bw_1^y)\right)^T g''_a(\bc_{mn}),\nonumber
\end{eqnarray}
where $\bc_{mn}=x_{mn}\bw_1^x+y_{mn}\bw_1^y+z_{mn}\bw_1^z+\bb_1$, and {indices $\kappa_1$ and $\kappa_2$ depend on $\kappa$}. The first and second derivatives of $g_a(\cdot)$ are
\begin{eqnarray}
g'_a(\cdot)=1-g^2_a(\cdot), \ \ \ 
g''_a(\cdot)=2g_a(\cdot)\left(g_a^2(\cdot)-1\right).\nonumber
\end{eqnarray} 
The CRLB of $\bp$ is given as
$\mathrm{CRLB}_{\bp}={\mathrm{Trace}(\boldsymbol{\mathcal{F}}^{-1}({\bp}))}$. 

\bibliographystyle{IEEEtran}
\bibliography{bibliography}

\end{document}

%% file: mydefinitions1.tex
\newcommand{\ba}{\boldsymbol{a}}

\newcommand{\bb}{\boldsymbol{b}}
\newcommand{\bc}{\boldsymbol{c}}

\newcommand{\bx}{\boldsymbol{x}}
\newcommand{\by}{\boldsymbol{y}}
\newcommand{\bh}{\boldsymbol{h}}
\newcommand{\bw}{\boldsymbol{w}}

\newcommand{\bp}{\boldsymbol{p}}

\newcommand{\br}{\boldsymbol{r}}

\newcommand{\bE}{\boldsymbol{E}}
\newcommand{\bP}{\boldsymbol{P}}
\newcommand{\bY}{\boldsymbol{Y}}
\newcommand{\bA}{\boldsymbol{A}}
\newcommand{\bB}{\boldsymbol{B}}
\newcommand{\bR}{\boldsymbol{R}}
\newcommand{\bZ}{\boldsymbol{Z}}
\newcommand{\bW}{\boldsymbol{W}}
\newcommand{\bI}{\boldsymbol{I}}
\newcommand{\bF}{\boldsymbol{F}}
\newcommand{\bC}{\boldsymbol{C}}

\newcommand{\bU}{\boldsymbol{U}}
\newcommand{\bV}{\boldsymbol{V}}
\newcommand{\bQ}{\boldsymbol{Q}}
\newcommand{\bH}{\boldsymbol{H}}
\newcommand{\bS}{\boldsymbol{S}}
\newcommand{\bG}{\boldsymbol{G}}
\newcommand{\bJ}{\boldsymbol{J}}

\newcommand{\bLambda}{\boldsymbol{\Lambda}}

\newcommand{\N}{\mathcal{N}}

\newcommand{\cev}[1]{\reflectbox{\ensuremath{\vec{\reflectbox{\ensuremath{#1}}}}}}

\newcommand{\herm}{^{\textrm{H}}}
\newcommand{\tra}{^{\textrm{T}}}

\newcommand{\CN}{\mathcal{CN}}

\newcommand{\bPhi}{\boldsymbol{\Phi}}

\usetikzlibrary{arrows,positioning,fit,shapes,calc,decorations.pathreplacing,matrix,patterns}

\newgeometry{left=1.5cm, right=1.5cm, top=1.5cm, bottom=1cm}
\tikzstyle{factornode} = [draw, fill=white, circle, inner sep=1pt, minimum size=1.4cm]
\tikzstyle{funnode} = [draw, rectangle,fill=black!100, minimum size = 8mm]